\documentclass[a4paper,11pt]{article}

\usepackage{jheppub}
\usepackage[T1]{fontenc}
\usepackage{braket}
\usepackage{cleveref}
\usepackage{xcolor}
\usepackage{multirow}
\usepackage{makecell}
\usepackage{amsmath}
\usepackage{amsfonts}
\usepackage{graphicx}
\usepackage{placeins}
\usepackage{bm}
\usepackage{slashed}
\usepackage{orcidlink}
\usepackage[font=small, labelfont=bf, textfont={small,it}]{caption}

\graphicspath{{figs/}}

\newcommand{\bee}{\begin{equation}}
\newcommand{\eee}{\end{equation}}
\newcommand{\beq}{\begin{eqnarray}}
\newcommand{\eeq}{\end{eqnarray}}
\newcommand{\beqnn}{\begin{eqnarray*}}
\newcommand{\eeqnn}{\end{eqnarray*}}

\newcommand{\sst}[1]{{\scriptscriptstyle{#1}}}
\newcommand{\sstrm}[1]{{\scriptscriptstyle{\rm #1}}}

\newcommand{\Tr}{\mathrm{Tr}}
\newcommand{\dd}{\mathrm{d}}
\newcommand{\ee}{\mathrm{e}}
\newcommand{\ii}{\mathrm{i}}
\newcommand{\U}{\mathrm{U}}
\newcommand{\SU}{\mathrm{SU}}
\newcommand{\etap}{\eta^{\prime}}

\newcommand{\chir}{\sst{(\chi)}}
\newcommand{\gtwo}{\tilde{g}_2}
\newcommand{\gtwop}{\tilde{g}^\prime_2}

\newcommand{\chiPT}{\chi\mathrm{PT}}

\newcommand{\Nf}{N_{\sstrm{f}}}
\newcommand{\Lpi}{\ell_{\pi\pi}}
\newcommand{\lSS}{\ell_{\scriptscriptstyle{\rm SS}}}
\newcommand{\lAA}{\ell_{\scriptscriptstyle{\rm AA}}}
\newcommand{\YMsup}{{\scriptscriptstyle{\rm (YM)}}}

\newcommand{\tildeS}{{\widetilde{S}}}

\newcommand{\TEK}{{\sstrm{TEK}}}

\newcommand{\s}{{\sstrm{s}}}
\newcommand{\R}{{\sstrm{R}}}
\newcommand{\LL}{{\sstrm{L}}}

\begin{document}
\title{\Huge Large-N $\bm{\pi^+\pi^+}$ scattering from twisted Eguchi--Kawai reduction}

\author[a]{Claudio Bonanno\,\orcidlink{0000-0002-1421-6510},}
\emailAdd{claudio.bonanno@unibe.ch}
\author[b]{Margarita Garc\'ia P\'erez\,\orcidlink{0000-0001-6566-2687},}
\emailAdd{margarita.garcia@csic.es}
\author[b,c]{Antonio Gonz\'alez-Arroyo\,\orcidlink{0000-0002-5263-1088},}
\emailAdd{A.Gonzalez-Arroyo@pm.me}
\author[d,e]{\\Ken-Ichi Ishikawa\,\orcidlink{0000-0001-6425-1894},}
\emailAdd{ishikawa@theo.phys.sci.hiroshima-u.ac.jp}
\author[e]{Masanori Okawa\,\orcidlink{0009-0007-7983-4364},}
\emailAdd{okawa@hiroshima-u.ac.jp}
\author[a]{Fernando Romero-L\'opez\,\orcidlink{0000-0003-0239-1394}.}
\emailAdd{fernando.romero-lopez@unibe.ch}

\affiliation[a]{Albert Einstein Center for Fundamental Physics, Institute for Theoretical Physics, University of Bern, Sidlerstra{\ss}e 5, CH-3012 Bern, Switzerland}

\affiliation[b]{Instituto de F\'isica Te\'orica UAM-CSIC, Calle Nicol\'as Cabrera 13-15,\\Universidad Aut\'onoma de Madrid, Cantoblanco, E-28049 Madrid, Spain}

\affiliation[c]{Departamento de F\'isica Te\'orica, Universidad Aut\'onoma de Madrid,\\M\'odulo 15, Cantoblanco, E-28049 Madrid, Spain}

\affiliation[d]{Core of Research for the Energetic Universe,\\Graduate School of Advanced Science and Engineering,\\Hiroshima University, Higashi-Hiroshima, Hiroshima 739-8526, Japan}

\affiliation[e]{Graduate School of Advanced Science and Engineering, Hiroshima University,\\Higashi-Hiroshima, Hiroshima 739-8526, Japan}

\preprint{}

\abstract{We present the first computation of the $\pi^+\pi^+$ scattering length in large-$N$ QCD in the 't Hooft limit, $\Nf/N\to 0$, using the twisted Eguchi--Kawai formulation of the lattice theory. We employ values of $N$ up to 1156 and consider several quark masses and lattice spacings to achieve controlled large-$N$, continuum, and chiral limits. The chiral behavior of the scattering length is compared with next-to-leading-order Chiral Perturbation Theory predictions. Our result for the low-energy coupling describing the sub-leading quark-mass dependence is consistent with existing S-matrix bootstrap bounds. Combining TEK results with previous lattice QCD data obtained using standard methods for $3\le N\le 6$, we are also able to extract sub-leading $\mathcal{O}(1/N)$ corrections to the large-$N$ scattering length.}


\maketitle
\flushbottom

\section{Introduction}

Pion-pion scattering is the simplest low-energy scattering process governed by strong interactions, while at the same time providing a genuinely non-perturbative probe of Quantum Chromo-Dynamics (QCD). Understanding and describing pion dynamics is a major goal of experimental investigations~\cite{Hyams:1973zf,Cohen:1973aaa,Losty:1974aaa,Durusoy:1973aaa,Hyams:1975aaa,Hoogland:1977aaa,Kaminski:1996da}, and has been the object of intense theoretical studies with a variety of methods: chiral effective theories and dispersive approaches~\cite{Dobado:1992ha,Bijnens:1995yn,Dobado:1996ps,Ananthanarayan:2000ht,Colangelo:2000jc,Colangelo:2001df,Caprini:2003ta,Pelaez:2004vs,Kaminski:2006yv,Pelaez:2006nj,GomezNicola:2007qj,Caprini:2011ky,Pelaez:2019eqa,Pelaez:2024uav,Colangelo:2025iuq,Colangelo:2026rvl}, lattice QCD calculations~\cite{Feng:2009ij,Dudek:2010ew,Fu:2011bz,NPLQCD:2011htk,Dudek:2012gj,Sasaki:2013vxa,Fu:2013ffa,ETM:2015bzg,Briceno:2016mjc,Briceno:2017max,Alexandrou:2017mpi,Mai:2019pqr,RBC:2021acc,RBC:2023xqv,Dawid:2025zxc,Baeza-Ballesteros:2022azb,Rodas:2023gma,DeGrand:2024lvp,Baeza-Ballesteros:2025iee,Boyle:2024grr}, and more recently, the S-matrix bootstrap program~\cite{Albert:2022oes,Fernandez:2022kzi,Ma:2023vgc,Albert:2023jtd,Albert:2023seb,Li:2023qzs,Guerrieri:2024jkn,Albert:2026xyz,Cheung:2026lpv,Bocchia:2026kew,Bocchia:2026lty}.

Pion-pion scattering has been also investigated in the 't Hooft limit of a large number of colors (large-$N$ limit)~\cite{tHooft:1973alw}, a general framework to study non-perturbative gauge dynamics. In this limit, achieved taking $N\to\infty$ at fixed 't Hooft coupling $\lambda=g^2N$ and number of flavors $\Nf$, expectation values can be expanded in a power series in $1/N$, which is fully non-perturbative in $\lambda$. Moreover, large-$N$ QCD exhibits important simplifications while retaining many of the essential features of the $N=3$ theory, prominently chiral symmetry breaking, confinement and asymptotic freedom. A relevant question related to large-$N$ $\pi\pi$ scattering is the fate of hadronic resonances as $N$ is varied, i.e., the dependence of their masses, widths, and trilinear couplings on $N$. This includes both conventional states, such as the $\rho$ resonance~\cite{Jenkins:1995vb}, as well as non-ordinary mesons~\cite{Cirigliano:2003yq}, or exotic candidates, such as tetraquarks~\cite{Weinberg:2013cfa,Knecht:2013yqa,Cohen:2014via,Allaman:2024vwn}. The scalar sector is particularly interesting in this regard, with the large-$N$ behavior of the $\sigma$ resonance being an active subject of investigation~\cite{Cohen:2014vta,Pelaez:2015qba,Afonin:2016evz,Pelaez:2021dak}. 

These topics are not only of theoretical interest, but have also phenomenological implications for strong interactions. For example, the large-$N$ behavior of the poles of non-ordinary mesons or tetraquark candidate states in scattering amplitudes can give information on their constituent composition in the $N=3$ theory~\cite{Cohen:2014vta,Pelaez:2015qba,Afonin:2016evz,Pelaez:2021dak, Weinberg:2013cfa,Knecht:2013yqa,Cohen:2014via,Allaman:2024vwn}. For these reasons, in recent years large-$N$ meson dynamics has been the object of several lattice studies~\cite{Bali:2013kia,DeGrand:2016pur,Hernandez:2019qed,Baeza-Ballesteros:2022azb,Donini:2020qfu,DeGrand:2020utq,DeGrand:2023hzz,DeGrand:2024frm,DeGrand:2024lvp,Baeza-Ballesteros:2025iee}, see also Refs.~\cite{GarciaPerez:2020gnf,Hernandez:2020tbc} for recent reviews. Indeed, lattice calculations are a tool that is able to provide unique first-principles predictions, as well as fundamental inputs for other complementary non-perturbative approaches, see, e.g., Refs.~\cite{Albert:2026xyz,Cheung:2026lpv}.

Motivated by these recent developments, the goal of this paper is to initiate a new research program targeting the lattice study of meson scattering using the Twisted Eguchi--Kawai (TEK) formulation of large-$N$ QCD~\cite{Gonzalez-Arroyo:1982hwr,PhysRevD.27.2397,Gonzalez-Arroyo:2010omx}. The TEK reduced matrix model provides a practical framework for the numerical resolution of QCD directly in the large-$N$ limit. It is based on the concept of large-$N$ volume independence~\cite{PhysRevLett.48.1063,BHANOT198247,Gross:1982at,Gonzalez-Arroyo:1982hwr,PhysRevD.27.2397,Aldazabal:1983ec,Kiskis:2002gr,Narayanan:2003fc,Kovtun:2007py,Unsal:2008ch,Gonzalez-Arroyo:2010omx,Neuberger:2020wpx} --- first established in the pioneering paper by Eguchi and Kawai~\cite{PhysRevLett.48.1063} --- and consists in performing lattice Monte Carlo calculations on a single-site lattice~\cite{Gonzalez-Arroyo:1983cyv,Das:1984jh,Das:1985xc,Narayanan:2004cp,Gonzalez-Arroyo:2005dgf,Kiskis:2009rf,Hietanen:2009ex,Hietanen:2010fx,Bringoltz:2011by,Hietanen:2012ma,Gonzalez-Arroyo:2012euf,Gonzalez-Arroyo:2013bta,Lohmayer:2013spa,Gonzalez-Arroyo:2014dua,GarciaPerez:2014azn,GarciaPerez:2015rda,Perez:2015ssa,Gonzalez-Arroyo:2015bya,Perez:2017jyq,GarciaPerez:2020gnf,Perez:2020vbn,Butti:2023hfp,Bonanno:2023ypf,Butti:2022sgy,Bonanno:2024bqg,Bonanno:2024onr,Bonanno:2025hzr} with twisted boundary conditions.\footnote{Twisted boundary conditions are necessary to avoid spontaneous center symmetry breaking on the one-point box (a crucial assumption for large-$N$ reduction). Other strategies can also be employed to this end, see, e.g., Refs.~\cite{BHANOT198247,Narayanan:2003fc,Kovtun:2007py,Unsal:2008ch,Neuberger:2020wpx}.} Getting rid of space-time degrees of freedom allows to reach much larger values of $N$ compared to the standard approach, while keeping the computational cost of the simulation under control.

Large-$N$ volume independence guarantees that no physical information is lost by reducing the lattice to one point at infinite $N$. To better clarify this statement, it is useful to recall that the finite-$N$ corrections of the model are of a different nature than the ones of the ordinary theory. They are equivalent to those of a lattice version of a non-commutative field theory formulated on a non-commutative torus of linear size $\sqrt{N}$~\cite{Gonzalez-Arroyo:1983zev,Eguchi:1982ta,Connes:1997cr,Douglas:2001ba,Ambjorn:1999ts,Ambjorn:2000nb,Ambjorn:2000cs}. The planar sector is exactly equivalent to a theory formulated on a finite lattice of linear size $\sqrt{N}$. This implies that the large-$N$ limit is also an infinite-volume limit, and that leading finite-$N$ corrections in TEK take the form of finite volume effects. Additional sub-leading finite-$N$ corrections arise from the non-planar sector, and can be modulated and reduced by appropriate choices of the twist fluxes. The genuine $1/N$ corrections of the original theory can be obtained by combining TEK results (which provide the $N=\infty$ limit) with standard finite-$N$ ones. This fact highlights the complementarity between the two approaches.

Over the last few years, the TEK formulation has been successfully applied to a variety of gluonic and hadronic observables in large-$N$ QCD, including the string tension~\cite{Gonzalez-Arroyo:2012euf}, the running coupling~\cite{GarciaPerez:2014azn} and the $\Lambda$-parameter~\cite{Butti:2023hfp}, meson masses and decay constants~\cite{Gonzalez-Arroyo:2015bya,Perez:2020vbn,Bonanno:2025hzr}, and the chiral condensate~\cite{Bonanno:2023ypf,Bonanno:2025bla}. It has been also successfully applied to other gauge theories, such as $\mathcal{N}=1$ supersymmetric large-$N$ Yang--Mills~\cite{Butti:2022sgy,Bonanno:2024bqg,Bonanno:2024onr} and adjoint large-$N$ QCD~\cite{Gonzalez-Arroyo:2013bta,GarciaPerez:2015rda,Hamada:2025whg}. All of these studies have primarily focused on static quantities or observables involving single-hadron states. A natural next step is therefore to ask whether genuinely dynamical quantities, such as scattering amplitudes and hadronic couplings, can also be determined within the TEK framework.

In this work, we develop a strategy to study meson scattering directly in large-$N$ QCD using the TEK model inspired by the finite-volume formalism used in lattice QCD scattering calculations~\cite{Luscher:1986aaa,Luscher:1986bbb}. In the strict $N\to\infty$ limit, mesons become non-interacting, and their scattering amplitudes vanish. The relevant quantities are therefore the coefficients governing the leading terms in the $1/N$ expansion of the amplitude, which can be determined directly from suitable Euclidean correlation functions computed in the TEK framework. As a first application, we focus on pion-pion scattering in the isospin-2 channel ($\pi^+ \pi^+$), which provides a particularly clean setting for establishing the methodology.

By studying the quark-mass dependence of the scattering amplitude and comparing it with predictions from Chiral Perturbation Theory ($\chiPT$), we determine the corresponding large-$N$ behavior of combinations of the Gasser--Leutwyler low-energy constants~\cite{Gasser:1983yg,Gasser:1984gg} describing the sub-leading term in the chiral expansion of the $\pi^+ \pi^+$ scattering length. This provides a direct connection between the large-$N$ dynamics accessible through TEK and the low-energy effective description of large-$N$ QCD. Moreover, as earlier pointed out, such results are potentially interesting also for other approaches studying pion dynamics at large $N$, e.g. the S-matrix bootstrap program~\cite{Albert:2026xyz,Cheung:2026lpv}. Pion-pion scattering has also been studied using conventional dynamical-fermion lattice simulations for several finite values of the number of colors, $N=3,4,5,6$~\cite{Baeza-Ballesteros:2022azb,Baeza-Ballesteros:2025iee,DeGrand:2024lvp}. By combining those results with the new large-$N$ information obtained here, we achieve a significantly more precise determination of the $N\to\infty$ limit and, at the same time, we can constrain sub-leading corrections in the $1/N$ expansion more effectively.

This paper is organized as follows: in Sec.~\ref{sec:setup} we briefly summarize the TEK formulation of large-$N$ QCD on the lattice; in Sec.~\ref{sec:scat_TEK} we establish the methodology to extract scattering information in the TEK model; in Sec.~\ref{sec:details} we summarize the strategy and present the lattice setup; in Sec.~\ref{sec:results} we present our results for the $\pi^+\pi^+$ scattering length; finally, in Sec.~\ref{sec:conclu} we draw our conclusions and discuss future outlooks of this study.

\section{Twisted Eguchi--Kawai formulation of \texorpdfstring{large-$N$}{large-N} lattice QCD}\label{sec:setup}

In the 't Hooft large-$N$ limit, $N\to\infty$ with $\Nf/N\to 0$, QCD becomes a theory of dynamical gluons and quenched quarks. In practice, this means that the Dirac determinant can be omitted from the generating functional of the theory. Gluon fields are therefore sampled from the Monte Carlo according to the pure-glue path integral probability distribution. They are then ``fed'' to the discretized Dirac operator entering the computation of fermionic observables. In this way, quarks ``feel'' gluon dynamics without any back-reaction.

In the TEK formulation, the generating functional of large-$N$ QCD thus reads:
\beq\label{eq:part_func}
\mathcal{Z}_\TEK \equiv \int [\dd U]\, \exp\left\{-\mathcal{S}_{\TEK}^{\YMsup}[U]\right\}, \quad [\dd U] \equiv \left[\prod_{\mu\,=\,1}^{d}\dd U_\mu\right] \, .
\eeq
Here $d=4$, $[\dd U]$ is the $\SU(N)$ invariant Haar measure, and $\mathcal{S}_{\TEK}^{\YMsup}$ is the TEK pure-Yang--Mills action. In this study we adopt the Wilson plaquette discretization:
\beq\label{eq:TEK_Wilson_action}
\mathcal{S}_{\TEK}^{\YMsup}[U] = -N b \sum_{\mu\,=\,1}^{d}\sum_{\nu \,\ne\, \mu} z_{\nu\mu} \Tr\left\{U_\mu U_\nu U_\mu^\dagger U_\nu^\dagger\right\},
\eeq
where only $d=4$ gauge links $U_\mu\in\SU(N)$ appear in the action, and where $b=1/\lambda_0$ is the inverse of the bare 't Hooft coupling $\lambda_0 = Ng_0^2$ (in terms of the usual inverse gauge coupling, one has $2N^2b=\beta$). By virtue of asymptotic freedom, the continuum limit $a\to 0$ (with $a$ the lattice spacing) is achieved by taking $b\to\infty$. The details of the Monte Carlo sampling algorithm of the path integral distribution $\exp\left\{-\mathcal{S}^{\YMsup}_{\TEK}[U]\right\}/\mathcal{Z}_{\TEK}$ can be found in Ref.~\cite{Perez:2015ssa}.

The twist factor $z_{\nu \mu}$ in Eq.~\eqref{eq:TEK_Wilson_action}, used to implement twisted boundary conditions, is chosen to be a $N^{\text{th}}$-root of 1:
\beq\label{eq:twist_def}
z_{\nu\mu} = z_{\mu\nu}^* = \exp\left\{\frac{2\pi \ii}{N} n_{\nu\mu} \right\}.
\eeq
Many choices for the integer-valued antisymmetric twist tensor $n_{\nu \mu}$ are possible. Here we adopt the so-called \emph{symmetric} twist. The number of colors $N$ is taken to be a perfect square $N=L^2$, while the twist tensor is taken to be:
\beq
n_{\nu \mu} = - n_{\mu \nu} = k(L) L, \quad \text{($\nu>\mu$)}.
\eeq
The flux parameter $k(L)$, taken to be a co-prime integer with $L$, needs to be increased as a function of $N$ as the large-$N$ limit is approached. As explained in Refs.~\cite{Gonzalez-Arroyo:2010omx,Chamizo:2016msz,GarciaPerez:2018fkj}, this is instrumental to avoid center-symmetry breaking~\cite{Ishikawa:2003,Bietenholz:2006cz,Teper:2006sp,Azeyanagi:2007su}, a crucial assumption underlying large-$N$ volume reduction. In addition, there are suitable choices of the flux $k$ that allow to reduce non-planar finite-$N$ corrections~\cite{Perez:2017jyq,Bribian:2019ybc}. As explained in the introduction, the usage of twisted boundary conditions has the effect of effectively enlarging the physical volume in which fields propagate. With the symmetric twist, the effective lattice size is given by:
\beq
L = \sqrt{N}.
\eeq
This relation shows mathematically how, in the large-$N$ limit (and with the adopted twist choice) space-time and color degrees of freedom are intertwined due to the Eguchi--Kawai reduction, and how the rank of the gauge group provides an effective volume.

Concerning the fermion sector, the TEK formulation of the Dirac operator in the fundamental representation was put forward originally in Ref.~\cite{Gonzalez-Arroyo:2015bya}. In a nutshell, the idea is to let quarks propagate on an extended lattice $L^3 \times T$, with $L=\sqrt{N}$ and $T=\ell_0 \sqrt{N}$ (we will choose here $\ell_0 = 2$), and in the presence of a periodic potential obtained replicating the gauge links $L=\sqrt{N}$ times, much like the Bloch description of the propagation of electrons in a crystal. In the end, it was shown in Ref.~\cite{Gonzalez-Arroyo:2015bya} that this amounts to use the following discretized Dirac operator (in the Wilson formulation) for the purpose of evaluating meson correlators in momentum space:
\beq\label{eq:Dirac_TEK_mom}
\widetilde{D}_{\TEK}(p_\mu) = \frac{1}{2\kappa}\mathbb{I} - \frac{1}{2} \sum_{\mu \, =\, 1}^{d}\left[(\mathbb{I}+\gamma_\mu) \otimes \ee^{\ii a p_\mu}\mathcal{W}_\mu + (\mathbb{I}-\gamma_\mu)\otimes \ee^{-\ii a p_\mu}\mathcal{W}_\mu^\dagger \right],
\eeq
with $\mathbb{I}$ the identity, $\kappa$ the Wilson hopping parameter, and $\gamma_\mu$ the Dirac matrices. The $N^2 \times N^2$ matrix $\mathcal{W}_\mu$ reads:
\beq\label{eq:Wilson_TEK_mat}
\mathcal{W}_\mu = U_\mu \otimes \Gamma_\mu^*.
\eeq
Here, the $\SU(N)$ matrices $\Gamma_\mu$ (\emph{twist eaters}) are solutions of the following equation:
\beq\label{eq:master_eq_twist_eaters}
\Gamma_\mu \Gamma_\nu = z_{\nu\mu}^* \Gamma_\nu \Gamma_\mu,
\eeq
with $z_{\nu\mu}$ the twist factor introduced earlier. Finally, the lattice momenta appearing in Eq.~\eqref{eq:Dirac_TEK_mom} are discretized in terms of the inverse lattice size as:
\beq
a p_0 = \frac{2\pi m_0}{T}, \quad a p_i = \frac{2\pi m_i}{L}, \qquad m_0=0,1,\dots,T-1, \quad m_i = 0,1,\dots,L-1.
\eeq

\section{Computation of the scattering length in TEK}\label{sec:scat_TEK}

This section presents the methodology to compute the scattering length in the TEK model using the L\"uscher formalism, as well as the numerical techniques used to evaluate the necessary correlators to extract finite-volume energies.

\subsection{Scattering lengths from the L\"uscher finite-volume formalism}\label{sec:scatlength_Luscher}

Let us start by working in the continuum theory defined in a finite three-dimensional spatial cubic box with volume $\ell^3$. The L\"uscher formula relates the infinite-volume scattering length $a_0$ to the finite-volume energy shift $\Delta E\equiv E_{\pi\pi} - 2m_\pi$ of the interacting $\pi\pi$ system with respect to the non-interacting one:
\bee\label{eq:luscher_formula}
\Delta E = E_{\pi\pi} - 2m_\pi = \frac{4\pi a_0}{m_\pi \ell^3} \left[1 + \mathcal{O}\left(\frac{a_0}{\ell}\right)\right],
\eee
where Eq.~\eqref{eq:luscher_formula} only retains the leading polynomial finite-volume effects, while exponentially small ones~\cite{Bedaque:2006yi} have been neglected. Note that, for the scattering length, we use the convention $q \cot \delta_0 = -1/a_0$ (where $\delta_0$ is the s-wave phase shift and $q$ the center-of-mass momentum). Since the $\mathrm{I}=2$ channel is repulsive, one has $\Delta E > 0$; this, with our sign convention for the scattering length, implies $a_0>0$.

In the large-$N$ limit, mesons become free, stable and non-interacting particles, thus $E_{\pi\pi} \to 2m_\pi + \mathcal{O}(1/N)$. Therefore:
\bee
\Delta E \underset{N\to\infty}{\sim} \mathcal{O}\left(\frac{1}{N}\right) \qquad \implies \qquad a_0 \underset{N\to\infty}{\sim} \mathcal{O}\left(\frac{1}{N}\right).
\eee
Moreover, $\Delta E$ vanishes as $\sim 1/\ell^3$ in the large-volume limit. For this reason, in the following we will consider the rescaled energy shift:
\bee\label{eq:largeN_energy_shift_def}
\Delta \varepsilon \equiv N \ell^3 \Delta E \underset{\substack{\ell \,\to\, \infty\\N \,\to\, \infty}}{\longrightarrow} \frac{4\pi N a_0}{m_\pi} \sim\mathcal{O}(N^0, \ell^0),
\eee
which is finite both in the large-$N$ limit and in the infinite-volume limit. This is very convenient, as in the TEK formulation these two limits are achieved simultaneously taking $N\to\infty$, since $\ell \propto \sqrt{N}$. Note that, according to L\"uscher formula, corrections to Eq.~\eqref{eq:largeN_energy_shift_def} are expected to be suppressed as $a_0/\ell \sim 1/(N\ell)$.

\subsection{Quark-mass dependence of the scattering length from \texorpdfstring{$\chiPT$}{ChiPT}}

Predictions for the quark-mass dependence of the scattering length $a_0$ can be derived in $\chiPT$, with the expansion coefficients depending on the particular isospin channel considered (as stated in the introduction, in this paper we will just focus on the isospin-2 case). Given that we work in the large-$N$ limit, we will need the results obtained in $\U(\Nf)$ $\chiPT$~\cite{Baeza-Ballesteros:2022azb}, where a light $\etap$ is explicitly included in the effective theory. In the end, in the large-$N$ limit one obtains:
\bee\label{eq:a0_chiPT_expansion}
\frac{a_0}{m_\pi} = \frac{1}{16 \pi F^2}\left[1-\left(\Lpi - \frac{\bar{\ell}_3}{2}\right)\xi + \mathcal{O}(\xi^2)\right], \qquad \quad \xi = \left(\frac{m_\pi}{4\pi F}\right)^2.
\eee
In the expansion parameter $\xi$, $m_\pi$ stands for the pion mass at finite quark mass, while $F\sim\mathcal{O}(\sqrt{N})$ denotes the chiral limit of the pion decay constant, related to the one at non-zero quark mass $F_\pi$ via (we here adopt the ``$F_\pi=92$ MeV'' convention):
\bee\label{eq:Fpi_chiPT_expansion}
F_\pi = F [1 + \bar{\ell}_4 \xi + \mathcal{O}(\xi^2)],
\eee
with $\bar{\ell}_4\sim\mathcal{O}(N)$ one of the seven Gasser--Leutwyler next-to-leading-order (NLO) low-energy constants (LECs), parameterizing the pion-mass dependence of $F_\pi$ at $\mathcal{O}(\xi)$~\cite{Gasser:1983yg,Gasser:1984gg}. The quantity $\bar{\ell}_3\sim\mathcal{O}(N)$ is another NLO LEC parameterizing the first non-linear correction in the quark mass to the Gell-Mann--Oakes--Renner relation:
\bee\label{eq:mpi_chiPT_expansion}
m_\pi^2 = 2 B_{\R} m_\R\left[1-\frac{\bar{\ell}_3}{2} \frac{2 B_{\R} m_\R}{(4\pi F)^2} + \mathcal{O}(m_\R^2)\right], \qquad B_{\R} = \Sigma_\R/F^2,
\eee
with $\Sigma_\R = - \lim_{m_{\R}\to 0} \braket{\overline{\psi}\psi}_\R\sim\mathcal{O}(N)$ the renormalized chiral condensate, and $m_\R$ the renormalized quark mass (note that $B_\R m_\R$ is renormalization-group invariant). Finally, $\Lpi\sim\mathcal{O}(N)$ is yet another NLO LEC that parameterizes the genuine sub-leading pion mass-dependence of the scattering length. It can be expressed in terms of the more customary $\chiPT$ NLO LECs $\bar{\ell}_1,\bar{\ell}_2\sim\mathcal{O}(N)$ as~\cite{FlavourLatticeAveragingGroupFLAG:2021npn}:
\bee
\Lpi = \frac{4}{3}\left(\bar{\ell}_1 + 2\bar{\ell}_2\right).
\eee
Note that chiral logs do not appear in these expressions, as they are sub-leading in the $1/N$ expansion, and that all the $\bar{\ell}_i$ denote scheme- and scale-independent quantities. Since $F^2\sim\mathcal{O}(N)$, one recovers $a_0\sim\mathcal{O}(1/N)$ from Eq.~\eqref{eq:a0_chiPT_expansion}, in agreement with general theoretical expectations about large-$N$ QCD. Moreover, the large-$N$ scaling of $F$ also implies that the expansion parameter $\xi\sim\mathcal{O}(1/N)$, and since $\bar{\ell}_{i\,=\,1,2,3,4}\sim\mathcal{O}(N)$, the product $(\Lpi-\bar{\ell}_3/2) \, \xi \sim\mathcal{O}(N^0)$, implying that the sub-leading term in the chiral expansion survives the large-$N$ limit.

\subsection{Extraction of the \texorpdfstring{large-$N$}{large-N} energy shift}\label{sec:largeN_energy_shift_extraction}

\begin{figure}[!t]
\centering
\includegraphics[scale=0.38]{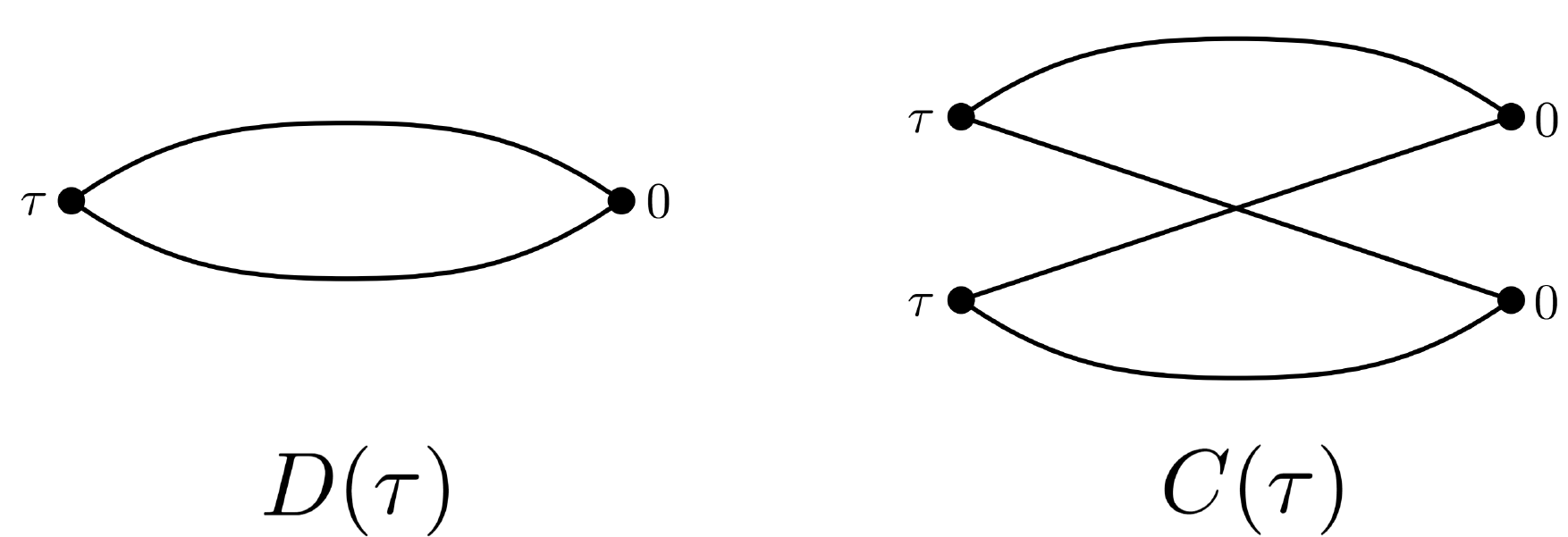}
\caption{Direct contraction $D(\tau)$ and crossed contraction $C(\tau)$ contributing to the $\pi^+$ and $\pi^+\pi^+$ correlators in Eqs.~\eqref{eq:Cpipi_Wick_contractions} and~\eqref{eq:Cpi_Wick_contractions}. Figure adapted from Ref.~\cite{Baeza-Ballesteros:2022azb}.}
\label{fig:pipi_I2_diagrams}
\end{figure}

Let us outline the general computational strategy to obtain the large-$N$ energy shift $\Delta \varepsilon$. We will start from a standard lattice theory defined on an extended space-time lattice with $L^3 \times T$ points. We will discuss later how to translate the relevant correlation functions in the reduced model.

Let us denote with $\mathcal{O}_\pi(\tau) \equiv \sum_{\vec{x}}\overline{d}(\vec{x},\tau)\gamma_5 u(\vec{x},\tau)$ and with $\mathcal{O}_{\pi\pi}(\tau)\equiv\mathcal{O}_\pi(\tau)\mathcal{O}_\pi(\tau)$ the $\pi^+$ and the $\pi^+\pi^+$ interpolating operators, respectively ($u$ and $d$ are the spinors of the up and down quarks respectively).
In the isospin-2 channel, performing the relevant Wick contractions one obtains:
\beq
\label{eq:Cpipi_Wick_contractions}
C_{\pi\pi}(\tau) = &\braket{\mathcal{O}_{\pi\pi}(\tau)\mathcal{O}_{\pi\pi}^\dagger(0)} &= \braket{2D^2(\tau)-2C(\tau)},\\
\label{eq:Cpi_Wick_contractions}
C_{\pi}(\tau) = &\braket{\mathcal{O}_\pi(\tau)\mathcal{O}_\pi^\dagger(0)} &= \braket{D(\tau)}
\eeq
The notation for the contractions is taken from Refs.~\cite{Gupta:1993rn,Liu:2009uw,Blum:2011pu}: $D$ stands for ``direct'' diagram, and is just the pion propagator between the time slices 0 and $\tau$, while $C$ is the ``crossed'' diagram. The diagrams are depicted in Fig.~\ref{fig:pipi_I2_diagrams}, and can be written as follows:
\bee\label{eq:C_trace}
C(\tau) = \sum_{\substack{\vec{x_1},\vec{y_1}\\\vec{x_2},\vec{y_2}}}\Tr\left[\gamma_5S(\vec{x}_1,\tau; \vec{y}_2,0)\gamma_5 S(\vec{y}_2,0;\vec{x}_2,\tau)\gamma_5S(\vec{x}_2,\tau; \vec{y}_1,0)\gamma_5S(\vec{y}_1,0;\vec{x}_1,\tau)\right],
\eee
\beq\label{eq:D_trace}
D(\tau) = \sum_{\vec{x}_1,\vec{y}_1}\Tr\left[\gamma_5S(\vec{x}_1,\tau; \vec{y}_1,0)\gamma_5 S(\vec{y}_1,0;\vec{x}_1,\tau)\right] .
\eeq
Here $S(x_\mu;y_\mu) \equiv \slashed{D}_{\LL}^{-1}(x_\mu;y_\mu)$ denotes the quark propagator from the space-time point $x_\mu$ to the space-time point $y_\mu$, given by the inverse of the lattice Dirac operator $\slashed{D}_{\LL}$.

For reasons that will become clear in the next section, it is also useful to write the corresponding expressions for $C(\tau)$ and $D(\tau)$ in momentum space. Introducing the quark propagator in momentum space $\tildeS(\vec{p},p_0)$,
\bee\label{eq:quark_prop_mom}
S(x_\mu;y_\mu) = \frac{1}{V_s\mathcal{T}}\sum_{p_0,\vec{p}} \ee^{-\ii p_\mu (x-y)_\mu} \tildeS(\vec{p},p_0), \qquad p_\mu=(\vec{p},p_0), \quad V_s=(aL)^3, \,\, \mathcal{T}=aT,
\eee
one gets for the $C$ contraction:
\bee
\begin{aligned}
C(\tau) = \sum_{\substack{\vec{x}_1,\vec{x}_2\\\vec{y}_1,\vec{y}_2}} \sum_{\substack{\vec{p},\vec{p}^{\,\prime}\\\vec{q},\vec{q}^{\,\prime}}} \sum_{\substack{p_0,p_0^\prime\\q_0,q_0^\prime}} & \ee^{-\ii \tau (p_0+q_0-q_0^\prime-p_0^\prime) -\ii\vec{p}\cdot(\vec{x}_1-\vec{y}_2) -\ii\vec{p}^{\,\prime}\cdot(\vec{y}_2-\vec{x}_2) -\ii\vec{q}\cdot(\vec{x}_2-\vec{y}_1) -\ii\vec{q}^{\,\prime}\cdot(\vec{y}_1-\vec{x}_1)}
\\
& \times \frac{1}{V_s^4\mathcal{T}^4} \Tr\left[\gamma_5\tildeS(\vec{p},p_0)\gamma_5\tildeS(\vec{p}^{\,\prime},p^\prime_0)\gamma_5\tildeS(\vec{q},q_0)\gamma_5\tildeS(\vec{q}^{\,\prime},q^\prime_0)\right].
\end{aligned}
\eee
Recalling that $\frac{1}{V_s} \sum_{\vec{x}} \ee^{-\ii \vec{x}\cdot(\vec{p}\,-\,\vec{q})} = \delta^{(3)}(\vec{p}-\vec{q})$, and using time-translation invariance, after some algebra one is finally left with:
\beq
\label{eq:C_tau_Fourier_1}
C(\tau) &=& \frac{1}{\mathcal{T}^3}\sum_{p_0,p_0^\prime,q_0} \ee^{-\ii\tau(p_0-p_0^\prime+q_0)} \widetilde{C}(p_0,p_0^\prime,q_0),\\
\label{eq:C_tau_Fourier_2}
\widetilde{C}(p_0,p_0^\prime,q_0) &=& \Tr\left[\gamma_5\tildeS(\vec{0},0)\gamma_5\tildeS(\vec{0},p_0)\gamma_5\tildeS(\vec{0},p^\prime_0)\gamma_5\tildeS(\vec{0},q_0)\right]
\eeq
A similar calculation for the $D$ contraction leads to:
\beq
\label{eq:D_tau_Fourier_1}
D(\tau) &=& \frac{1}{\mathcal{T}}\sum_{p_0} \ee^{-\ii \tau p_0} \widetilde{D}(p_0),\\
\label{eq:D_tau_Fourier_2}
\widetilde{D}(p_0) &=& \Tr\left[\gamma_5\tildeS(\vec{0},0)\gamma_5\tildeS(\vec{0},p_0)\right].
\eeq
Note that in all cases the factors of $1/\mathcal{T}$ match the number of momentum sums, implying a correct normalization of the correlators.

We are now in the position of discussing the calculation of $\Delta \varepsilon$. The starting point is the asymptotic behavior of the 1-pion and of the 2-pion correlation functions for large time separations:
\beq\label{eq:pipicorr_largetau_behavior}
C_{\pi\pi}(\tau) &\underset{\tau\to\infty}{\sim}& A_{\pi\pi}\,\ee^{-(2m_\pi+\Delta E)\tau},\\
C_{\pi}(\tau) &\underset{\tau\to\infty}{\sim}& A_{\pi}\,\ee^{-m_\pi\tau},
\eeq
In the large-$N$ limit, it is possible to relate the amplitudes $A_\pi$ and $A_{\pi\pi}$ invoking large-$N$ factorization:
\beq
\braket{D^2(\tau)} = \braket{D(\tau)}^2 [1 + \mathcal{O}(1/N^2)], \label{eq:factorization}
\eeq
i.e., the expectation value of a double trace $\braket{D^2(\tau)}$  becomes equal to the product of two single-trace expectation values $\braket{D(\tau)}^2$. Therefore, retaining only the leading order in the large-$N$ limit [recall that $\Delta E \sim \mathcal{O}(1/N)$]:
\beq\label{eq:pipicorr_largeN}
C_{\pi\pi}(\tau) = 2\braket{D(\tau)}^2 &\underset{\tau\to\infty}{\sim}& A_{\pi\pi}\ee^{-2m_\pi\tau} [1 + \mathcal{O}(1/N)],\\
&\underset{\tau\to\infty}{\sim}& 2 \left(A_{\pi}\ee^{-m_\pi\tau}\right)^2 [1 + \mathcal{O}(1/N)],
\eeq
i.e., $A_{\pi\pi}=2A_\pi^2$. Retaining now also sub-leading terms in the $1/N$ expansion, one has:
\beq
C_{\pi\pi}(\tau) &\underset{\tau\to\infty}{\sim}& A_{\pi\pi} \ee^{-(2m_\pi+\Delta E)\tau}\\
&=& A_{\pi\pi} \ee^{-2m_\pi\tau} \ee^{-\Delta E\tau}\\
\label{eq:1Nexp_2pi_corr}
&=& 2A^2_\pi \ee^{-2m_\pi\tau} \left[ 1 \,-\, (\Delta E \tau) \, + \, \mathcal{O}\left(\frac{1}{N^2}\right) \right]\\
&=& 2C_\pi^2(\tau) \left[ 1 \,-\, (\Delta E \tau) \, + \, \mathcal{O}\left(\frac{1}{N^2}\right) \right].
\eeq
This implies that the ratio of the 2-pion correlator $C_{\pi\pi}$ to the square of the 1-pion correlator $C_\pi^2$ has the following asymptotic behavior:
\beq
\frac{C_{\pi\pi}(\tau)}{C_\pi^2(\tau)} \underset{\tau\to\infty}{\sim} 2\left[ 1 \,-\, (\Delta E \tau) \, + \, \mathcal{O}\left(\frac{1}{N^2}\right) \right].
\eeq
This ratio can be expressed in terms of the contractions $D(\tau)$ and $C(\tau)$ as follows:
\begin{align}\label{eq:R2_derivation}
\begin{split}
\frac{C_{\pi\pi}(\tau)}{C_{\pi}^2(\tau)} &= \frac{2\braket{D^2(\tau)} -2\braket{C(\tau)} }{\braket{D(\tau)}^2}\\
&= \frac{2\braket{D(\tau)}^2 -2\braket{C(\tau)} }{\braket{D(\tau)}^2} + \mathcal{O}\left(\frac{1}{N^2}\right)\\
&= 2 - 2\frac{\braket{C(\tau)}}{\braket{D(\tau)}^2} + \mathcal{O}\left(\frac{1}{N^2}\right)\\
&= 2\left[1 - \frac{\braket{\overline{C}(\tau)}}{\braket{\overline{D}(\tau)}^2N \ell^3} + \mathcal{O}\left(\frac{1}{N^2}\right)\right] \underset{\tau\to\infty}{\sim} 2 \left[1-\Delta E \tau + \mathcal{O}\left(\frac{1}{N^2}\right)\right].
    \end{split}
\end{align}
In the second line of Eq.~\eqref{eq:R2_derivation} we have used Eq.~\eqref{eq:factorization}, while in the last equality we have used the following scaling properties of $\braket{D(\tau)}$ and $\braket{C(\tau)}$ with the number of colors $N$ and the spatial volume $V_s=\ell^3=(aL)^3$:
\beq
\braket{D(\tau)} \underset{\substack{\ell \,\to\,\infty\\N\,\to\,\infty}}{\sim} \braket{\overline{D}(\tau)} N \ell^3, \qquad \qquad
\braket{C(\tau)} \underset{\substack{\ell \,\to\,\infty\\N\,\to\,\infty}}{\sim} \braket{\overline{C}(\tau)} N \ell^3,
\eeq
where the bar denotes a quantity that is intensive with respect to both $N$ and $\ell$. These scaling behaviors follow from translational invariance and from the trace over color indices appearing in $C_{\pi}$ and $C_{\pi\pi}$. We thus finally obtain:
\beq\label{eq:ratio_R2_final}
\mathcal{R}_{2}(\tau) \equiv \frac{\braket{\overline{C}(\tau)}}{\braket{\overline{D}(\tau)}^2} \underset{\tau\to\infty}{\sim} (N \ell^3\Delta E) \tau = \Delta \varepsilon \tau.
\eeq
In Eq.~\eqref{eq:ratio_R2_final}, both the left-hand and the right-hand sides have a finite large-$N$ and infinite-volume limit. Indeed, the left-hand side is expressed in terms of intensive quantities in both $N$ and $\ell^3$, while the right-hand side, using the L\"uscher formula~\eqref{eq:luscher_formula}, is exactly the $\mathcal{O}(\ell^0)$ and $\mathcal{O}(N^0)$ combination in Eq.~\eqref{eq:largeN_energy_shift_def}. Therefore, this is a formula that can be employed in the TEK formulation, where the two limits are achieved simultaneously taking $N\to\infty$. In particular, our goal is to compute the ratio of correlators on the left-hand side of Eq.~\eqref{eq:ratio_R2_final} within the TEK formulation, and extract $\Delta \varepsilon$ from the linear slope of $\mathcal{R}_2(\tau)$ at sufficiently large Euclidean time separations.

\subsection{The 2-pion correlation function in the reduced model}\label{sec:2picorr_TEK}

In a standard extended lattice, one could directly evaluate Eq.~\eqref{eq:ratio_R2_final} by computing the contractions $C$ and $D$ from their expressions in Eqs.~\eqref{eq:C_trace} and~\eqref{eq:D_trace}. Those formulas, however, are not useful in the TEK formulation due to the space-time volume reduction. In order to compute $D(\tau)$, we will adopt the  strategy put forward in Ref.~\cite{Gonzalez-Arroyo:2015bya}, and already implemented for the calculation of 1-meson correlators in the previous TEK studies~\cite{Perez:2020vbn,Bonanno:2025hzr}, which can also be easily generalized to the 4-point function $C(\tau)$. As already anticipated in Sec.~\ref{sec:setup}, in~\cite{Gonzalez-Arroyo:2015bya} it was shown that one can evaluate meson correlation functions in momentum space using the Dirac operator in Eq.~\eqref{eq:Dirac_TEK_mom}. Then, one can obtain the ones in coordinate space by means of a Fourier transform. In momentum space, any problem related to the reduction of the space-time degrees of freedom is bypassed: momenta explicitly appear in the discretized TEK Dirac operator~\eqref{eq:Dirac_TEK_mom}, and are correctly discretized in terms of the inverse of the effective size. Therefore, the useful formulas on the reduced lattice are those in Eqs.~(\ref{eq:C_tau_Fourier_1}--\ref{eq:D_tau_Fourier_2}).

An important point to be taken into account when writing fermion correlators in the reduced model is that the full TEK Dirac operator assumes a block-diagonal structure, where each of the $\sqrt{N}$ blocks is given by the Dirac operator in Eq.~\eqref{eq:Dirac_TEK_mom}, see~\cite{Gonzalez-Arroyo:2015bya}. Therefore, any fermion correlator built in terms of $\widetilde{D}_{\TEK}^{-1}$ should be multiplied by a factor of $\sqrt{N}$ to take into account the full block-diagonal structure. This is particularly relevant for the purpose of computing the normalized contractions $\overline{C}(\tau)$ and $\overline{D}(\tau)$, as missing this factor of $\sqrt{N}$ would yield the wrong normalization. In the end, one has (in lattice units):
\beq
\label{eq:C_TEK}
\overline{C}(\tau) &=& \sqrt{N} \frac{1}{N L^3} \frac{1}{T^3}\sum_{p_0,p_0^\prime,q_0} \ee^{-\ii\tau(p_0-p_0^\prime+q_0)} \, \Tr\left[\gamma_5\tildeS(0)\gamma_5\tildeS(p_0)\gamma_5\tildeS(p^\prime_0)\gamma_5\tildeS(q_0)\right],\\
\label{eq:D_TEK}
\overline{D}(\tau) &=& \sqrt{N} \frac{1}{N L^3} \frac{1}{T}\sum_{p_0} \ee^{-\ii\tau p_0} \, \Tr\left[\gamma_5\tildeS(0)\gamma_5\tildeS(p_0)\right],
\eeq
where $L=\sqrt{N}$, $T=2\sqrt{N}$, and where from now on
\bee
\tildeS(p_0) \equiv \widetilde{D}_{\TEK}^{-1}(\vec{0},p_0)
\eee
will indicate the zero-spatial-momentum lattice quark propagator in the reduced model. Since the Dirac operator $\widetilde{D}_\TEK$ is a $N^2\times N^2$ matrix, cf.~Eqs.~\eqref{eq:Dirac_TEK_mom} and~\eqref{eq:Wilson_TEK_mat}, the factor of $\sqrt{N}/(NL^3)=1/N^2$ correctly normalizes the traces appearing in Eqs.~\eqref{eq:C_TEK} and~\eqref{eq:D_TEK}, thus giving the desired intensive quantities in $N$.

\subsection{Stochastic calculation of fermion traces}

In order to compute the 4-point contraction
\beq
\braket{\overline{C}(\tau)} = \frac{1}{N^2}\frac{1}{T^3} \sum_{p_0,p_0^\prime,q_0} \ee^{-\ii \tau (p_0 - p_0^\prime +q_0)} \, \braket{\widetilde{C}(p_0,p_0^\prime,q_0)},
\eeq
we will evaluate the following trace:
\beq\label{eq:initial_trace}
\widetilde{C}(p_0,p_0^\prime,q_0) \to \Tr\left[\gamma_5^{\ell_\s} \tildeS(0) \gamma_5^{\ell_\s} \tildeS(p_0) \gamma_5^{\ell_\s} \tildeS(p_0^\prime) \gamma_5^{\ell_\s} \tildeS(q_0)\right].
\eeq
Here, to improve the projection of the interpolating operator onto the 2-pion state, we have substituted $\gamma_5 \to \gamma_5^{\ell_\s}$, with $\gamma_5^{\ell_\s}$ indicating a smeared pseudo-scalar operator:
\beq
\gamma_{5}^{\ell_\s} \equiv D_{\s}^{\ell_\s} \gamma_{5}, \qquad D_{\s} \equiv \frac{1}{1+6c}\left[1+c \sum_{i\,=\,1}^{3}\left(\overline{U}_i\otimes\Gamma_i^{*}+\overline{U}^{\dagger}_i\otimes{\Gamma_i^{*}}^\dagger\right)\right].
\eeq
In this expression, the $\Gamma_i$ are the spatial twist-eaters, while $\overline{U}_i$ stands for the spatial links after $n_{\sstrm{APE}}$ APE smearing steps. The latter is an iterative smearing procedure where, at each step, the smeared link is related to the one at the previous one by:
\beq
U^{(n+1)}_i = \mathrm{Proj}_{\scriptscriptstyle{\SU(N)}}\left[(1-f)U^{(n)}_i + \frac{f}{4}\sum_{j \, \ne \, i}\left(z_{ij} U_j U_i U_j^\dagger + z_{ji} U_j^\dagger U_i U_j\right)\right],
\eeq
In practical calculations we used $n_{\scriptscriptstyle{\rm APE}}=10$, $f=0.081$, $c=0.4$. Concerning the level of smearing, to keep the number of inversions of the Dirac operator reasonable, we will choose a single suitable level $\ell_{\s}$ that is sufficiently high to allow us to neglect excited state contributions, but at the same time not excessively high to lose self-averaging. This optimal level of smearing can be chosen on the basis of the results of the previous TEK paper~\cite{Bonanno:2025hzr}, where a detailed study of the large-$N$ meson spectrum was performed. There, 1-meson correlation functions were obtained via the resolution of a Generalized Eigenvalue Problem (GEVP) involving a large operator basis obtained from considering several smearing levels $\ell_{\s}$. In the present study, we chose $\ell_{\s} = 50$: this was the largest smearing level used in the GEVP basis for the pion in~\cite{Bonanno:2025hzr}, and it showed a good suppression of excited states.

The calculation of the trace in~\eqref{eq:initial_trace} was done via stochastic sources:
\begin{align}
\widetilde{C}(p_0, p_0^\prime, q_0)
&\to \frac{1}{\mathcal{N}}\sum_{i\ne j} \text{tr}\left\{ z_i^\dagger \left[D_\s^{\frac{\ell_\s}{2}} \gamma_5 \tildeS(0)\gamma_5^{\ell_{\s}} \tildeS(p_0) D_\s^{\frac{\ell_\s}{2}} \right] z_j \, z_j^\dagger \left[D_\s^{\frac{\ell_\s}{2}} \gamma_5 \tildeS(p^\prime_0) \gamma_5^{\ell_\s} \tildeS(q_0) D_\s^{\frac{\ell_\s}{2}} \right] z_i \right\}
\nonumber\\
& \equiv \frac{1}{\mathcal{N}}\sum_{i\ne j} \text{tr}\left\{ \mathcal{M}_{ij}(0,p_0) \mathcal{M}_{ji}(p^\prime_0,q_0) \right\},
\label{eq:stoc_trace2}
\end{align}
Here, $\mathcal{N}=N_{\sstrm{src}}(N_{\sstrm{src}}-1)$ is a normalization factor, where $N_{\sstrm{src}}$ indicates the number of spin-diagonal random stochastic sources $z_i$ that are used to evaluate traces over color ($U_\mu$) and space-time ($\Gamma_\mu$) indices. For all calculations we have used $N_{\sstrm{src}} = 10$. The trace over Dirac indices, denoted with $\mathrm{tr}$, was instead computed exactly. Since $z_i$ are spin-diagonal, smearing operators are divided 
into two equal parts as $D_\s^{\ell_\s} \propto \sum_{i} D_\s^{\frac{\ell_\s}{2}} z_i z_i^\dagger D_\s^{\frac{\ell_\s}{2}}$ 
to reduce the computational cost of $\mathcal{M}_{ij}$. The quantity $\mathcal{M}_{ij}(P,Q)$ for two generic momenta $P$ and $Q$ can be rewritten as follows:
\begin{align}
\mathcal{M}_{ij}(P,Q) 
&= z_i^\dagger \left[D_\s^{\frac{\ell_\s}{2}} \gamma_5 \tildeS(P) D_\s^{\ell_\s} \gamma_5 \tildeS(Q) D_\s^{\frac{\ell_\s}{2}} \right] z_j
= z_i^\dagger \left[D_\s^{\frac{\ell_\s}{2}} \gamma_5 \tildeS(P) \gamma_5 D_\s^{\frac{\ell_\s}{2}} D_\s^{\frac{\ell_\s}{2}} \tildeS(Q) D_\s^{\frac{\ell_\s}{2}} \right] z_j
\nonumber\\
&= z_i^\dagger \left[D_\s^{\frac{\ell_\s}{2}} \tildeS(P)^\dagger D_\s^{\frac{\ell_\s}{2}} D_\s^{\frac{\ell_\s}{2}} \tildeS(Q) D_\s^{\frac{\ell_\s}{2}} \right] z_j = \left[\tildeS^{\sstrm{(s)}}_i(P)\right]^\dagger \left[\tildeS^{\sstrm{(s)}}_j(Q)\right],
\label{eq:stoc_final}
\end{align}
where we have introduced the following short-hand,
\beq
\label{eq:source}
\tildeS^{\sstrm{(s)}}_i(P) = D_\s^{\frac{\ell_\s}{2}} \tildeS(P) D_\s^{\frac{\ell_\s}{2}} z_i.
\eeq
Thus, solving Eq.~\eqref{eq:source} for all momenta and spin-diagonal stochastic sources $z_i$, one can compute $\mathcal{M}_{ij}$, cf.~Eq.~\eqref{eq:stoc_final}, and finally $\widetilde{C}$, cf.~Eq.~\eqref{eq:stoc_trace2}. This requires $T=2\sqrt{N}$ (for momenta $P$) times $4$ (for spin) for each $z_i$, thus a total of $ 4 T N_{\sstrm{src}} = 8 \sqrt{N} N_{\sstrm{src}}$ inversions.\\
The calculation of the trace in Eq.~\eqref{eq:stoc_final} with equal indices $i=j$ allows instead to compute the trace defining the single-pion correlation function, which is also an ingredient entering the calculation of the $\pi\pi$ energy shift. As a matter of fact, since
\beq
\braket{\overline{D}(\tau)} = \frac{1}{N^2}\frac{1}{T} \sum_{p_0}\ee^{-\ii \tau p_0} \braket{\widetilde{D}(p_0)}, \qquad \widetilde{D}(p_0) \to \Tr\left[\gamma_5^{\ell_\s} \tildeS(0) \gamma_5^{\ell_\s} \tildeS(p_0)\right].
\eeq
the stochastic computation of the $\widetilde{D}(p_0)$ can be performed via:
\beq
\widetilde{D}(p_0) \rightarrow \frac{1}{N_{\sstrm{src}}} \sum_i \mathrm{tr} \left\{\mathcal{M}_{ii}(0,p_0)\right\},
\eeq
with $\mathcal{M}_{ii}(0,p_0)$ the same quantity in Eq.~\eqref{eq:stoc_final} with $i=j$.

\section{Lattice strategy and simulation details}\label{sec:details}

In this section we recapitulate our strategy and summarize the employed lattice setup.

\subsection{Summary of the strategy}
Given that the chiral behavior of the pion decay constant down to the chiral limit was already computed (in the continuum limit) in~\cite{Bonanno:2025hzr} from the TEK model, the plan of the present paper is the following:
\begin{enumerate}
\item Extract the $\mathrm{I}=2$ energy shift $\Delta \varepsilon$ from the ratio of correlators in Eq.~\eqref{eq:ratio_R2_final}.
\item Building on the L{\"u}scher formula, extrapolate $\Delta \varepsilon$ towards the large-$N$ limit to extract $4 \pi N a_0/m_\pi$.
\item Compute the pion-mass dependence of $a_0/m_\pi$ in the continuum limit from a global chiral-continuum fit of lattice data.
\item Check that the chiral limit of $a_0/m_\pi$ is $1/(16 \pi F^2)$, as predicted by $\chiPT$. This is a non-trivial consistency test of the previous TEK calculation of the chiral-continuum limit of the large-$N$ pion decay constant $F/\sqrt{N}$ in Ref.~\cite{Bonanno:2025hzr}.
\item Extract the large-$N$ value of $\Lpi/N$ from the slope of the $\xi$-dependence of $a_0/m_\pi$. This is a \emph{novel prediction} of this study. To this end, one also needs $\bar{\ell}_3/N$, which can be extracted from the pion mass data as a function of the quark mass already presented in Ref.~\cite{Bonanno:2025hzr}. This calculation is discussed in Appendix~\ref{app:l3}. Our result for $\Lpi$ puts a non-perturbative bound on the values of $\bar{\ell}_1$ and $\bar{\ell}_2$, which will be compared with existing bootstrap ones on these couplings.
\item Combine the large-$N$ TEK result for $\Lpi/N$ with the finite-$N$ ones given in~\cite{Baeza-Ballesteros:2022azb} to compute the $\mathcal{O}(1/N)$ corrections to this quantity. Given that the $\mathcal{O}(1/N)$ corrections to $F/\sqrt{N}$ were already computed in~\cite{Bonanno:2025hzr}, in the end we will provide NLO results for the isospin-2 $\pi\pi$ scattering length both in the chiral and in the $1/N$ expansions.
\end{enumerate}

\begin{table}[!t]
\small
\begin{center}
\begin{tabular}{|c|c|c|c|c|c|c|c|}
\hline
$b$ & $a\sqrt{\sigma}$ & $a/\sqrt{8t_1}$ & $N$ & $k$ & $\kappa$ & \makecell{$a m_\pi$\\(this work)} & \makecell{$am_\pi$\\\cite{Bonanno:2025hzr}} \\
\hline
\multirow{5}{*}{0.355} & \multirow{5}{*}{0.2410(30)} & \multirow{5}{*}{0.37707(57)} & \multirow{5}{*}{\makecell{289, 361, 529,\\841, 1156}} & \multirow{5}{*}{\makecell{5, 7, 7,\\9, 13}} & 0.1575 & 0.5277(67) & 0.5272(39) \\
& & & & & 0.1585 & 0.4821(36) & 0.4859(30)\\
& & & & & 0.1592 & 0.4367(43) & 0.4363(33)\\
& & & & & 0.1600 & 0.3904(28) & 0.3834(31)\\
& & & & & 0.1607 & 0.3407(28) & 0.3367(34)\\
\hline
\multirow{4}{*}{0.360} & \multirow{4}{*}{0.2058(25)} & \multirow{4}{*}{0.31776(50)} & \multirow{4}{*}{\makecell{289, 361\\529, 841}} & \multirow{4}{*}{\makecell{5, 7\\7, 9}} & 0.1560 & 0.4550(46) & 0.4547(71)\\
& & & & & 0.1570 & 0.3947(16) & 0.3834(44)\\
& & & & & 0.1577 & 0.3490(16) & 0.3467(51)\\
& & & & & 0.1585 & 0.2887(40) & 0.2765(20)\\
\hline
0.365 & 0.1783(17) & 0.27181(74) & 841 & 9 & 0.1567 & 0.2468(43) & \multicolumn{1}{c}{} \\
\cline{1-7}
0.370 & 0.1573(19) & 0.23607(61) & 841 & 9 & 0.1543 & 0.2859(33) & \multicolumn{1}{c}{}\\
\hline
0.375 & 0.1361(17) & 0.20576(55) & 841 & 9 & 0.1527 & 0.2997(47) & 0.2862(27)\\
\hline
\end{tabular}
\end{center}
\caption{Summary of the simulation parameters of the ensembles employed in this study. Scale setting quantities come from Ref.~\cite{Gonzalez-Arroyo:2012euf} (string tension $\sigma$) and Ref.~\cite{Butti:2023hfp,Bonanno:2025hzr} (reference gradient flow scale $t_1$). Pion masses have been obtained anew from the plateaus in the effective mass of $\braket{\overline{D}(\tau)}$. For comparison, we also report the GEVP pion mass results of Ref.~\cite{Bonanno:2025hzr}.}
\label{tab:simulation_points}
\end{table}

\subsection{Simulation details}

In this work, we use a variety of lattice parameters $(b,\kappa,N)$ in order to control the continuum, chiral, and infinite-volume extrapolations, summarized in Tab.~\ref{tab:simulation_points}.

Let us explain the reasoning behind their selection. For two values of the bare coupling, $b=0.355$ and $b=0.360$, corresponding to the two coarsest lattice spacings, we considered several values of the quark mass in order to control the chiral extrapolation. Moreover, by having several values of $N$, which determines the effective volume, we can safely get rid of finite-volume effects for each $(b,\kappa)$ combination. For these 41 combinations, this leads to a robust determination of the chiral behavior of $a_0$ for two values of the lattice spacing.

We also considered three further ensembles at finer lattice spacings, corresponding to bare couplings $b=0.365, 0.370, 0.375$. For these, we only considered one value of the number of colors, $N=841$, and one value of the quark mass each. The quark masses for these ensembles were chosen to be evenly distributed across the pion-mass range explored on the coarser $b=0.355,0.360$ ensembles. The role of these three ensembles is to provide further leverage on the continuum extrapolation. However, given that only one value of $N$ is available, the extrapolation of the scattering length to infinite effective volume must use an estimate of finite-$N$ effects obtained from the other two values of $b=0.355,0.360$.

The absolute scale was fixed via the gradient flow~\cite{Narayanan:2006rf,Luscher:2010iy} scale $t_1$, defined as:
\beq
\frac{1}{N}\braket{t^2E(t)}\bigg\vert_{t \, = \, t_1} = 0.05,
\eeq
with $E$ the clover action density computed at positive flow time $t>0$. This choice is motivated by its higher accuracy compared to other scale setting quantities at our disposal, e.g., the string tension $\sigma$. Conversion to string tension units is only performed at the very end to compare with previous results using the conversion factor determined in Ref.~\cite{Bonanno:2025hzr} in the continuum limit:
\beq\label{eq:t1_sigma_conv}
\sqrt{8t_1\sigma} = 0.6759(92).
\eeq

\section{Results}\label{sec:results}

In this section we present our results for the $\pi^+\pi^+$ scattering length in the large-$N$ limit. 

\subsection{Energy shift and scattering length calculation}

\begin{figure}[!t]
\centering
\includegraphics[scale=0.62]{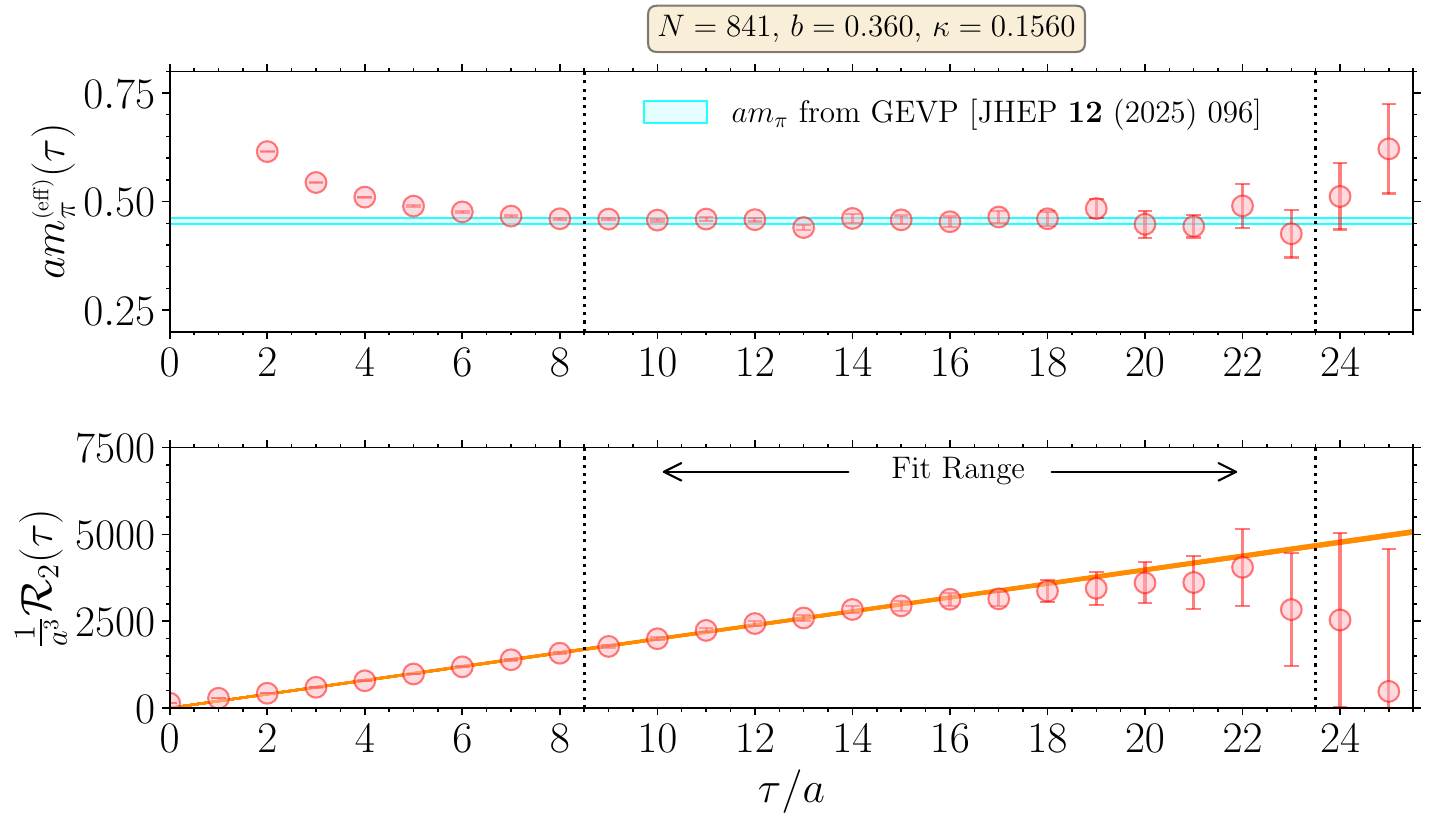}
\caption{Example of energy shift calculation from the slope of the linear large-time behavior of $\mathcal{R}_2(\tau)$. The linear fit to extract the slope is always performed in the range $\tau_{\min}/a=9$ and $\tau_{\max}/a=\sqrt{N}-6$ (vertical dashed lines), and errors are estimated via a binned jack-knife analysis. In all cases, this interval includes the range where the pion effective mass computed from $\braket{\overline{D}(\tau)}$ exhibits a plateau. This effective mass plateau is compared with the result obtained for $m_\pi$ from the more extensive GEVP analysis performed in~\cite{Bonanno:2025hzr} (shown as a horizontal shaded band).}
\label{fig:energy_shift_ex}
\end{figure}

The calculation of the energy shift is exemplified in Fig.~\ref{fig:energy_shift_ex}, and further plots are reported in Appendix~\ref{app:additional_plot}. As it can be seen, the ratio of correlators $\mathcal{R}_2(\tau)$ indeed exhibits a linear growth as a function of the time separation $\tau$, as expected from the general arguments discussed in Sec.~\ref{sec:largeN_energy_shift_extraction}. A linear fit to $\mathcal{R}_2(\tau)$ in the range where the effective pion mass built from $\braket{\overline{D}(\tau)}$ exhibits a plateau allows to determine the rescaled energy shift $\Delta \varepsilon$ without excited state contamination. All obtained results for $\Delta \varepsilon$ in lattice units are reported in Appendix~\ref{app:raw_data}. From the plateau in the effective mass of $\braket{\overline{D}(\tau)}$ we also extracted the values of the pion mass anew, finding agreement with the previous determinations of Ref.~\cite{Bonanno:2025hzr} in all cases (observed differences are 2.5 standard deviations at most).

\begin{figure}[!t]
\centering
\includegraphics[scale=0.5]{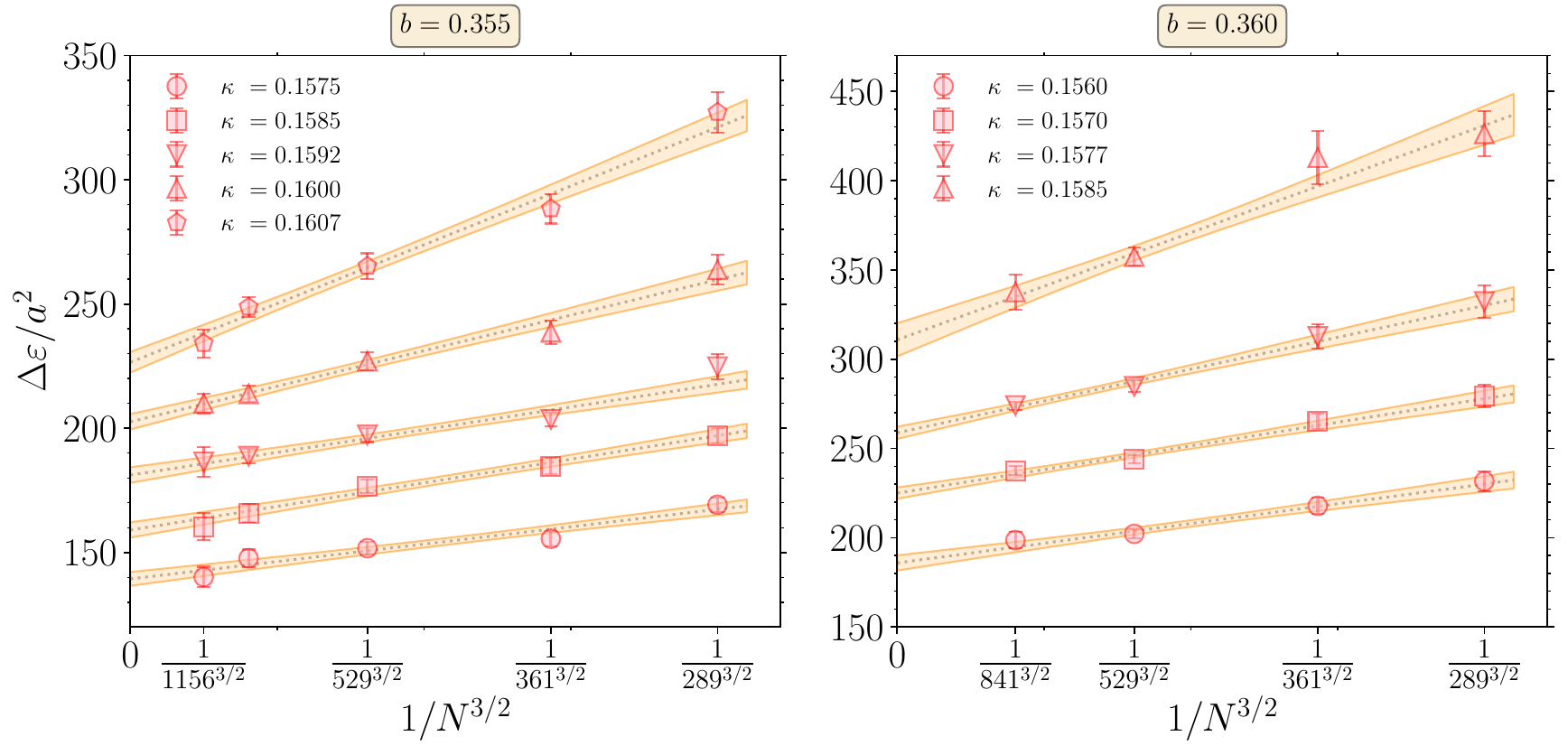}
\includegraphics[scale=0.5]{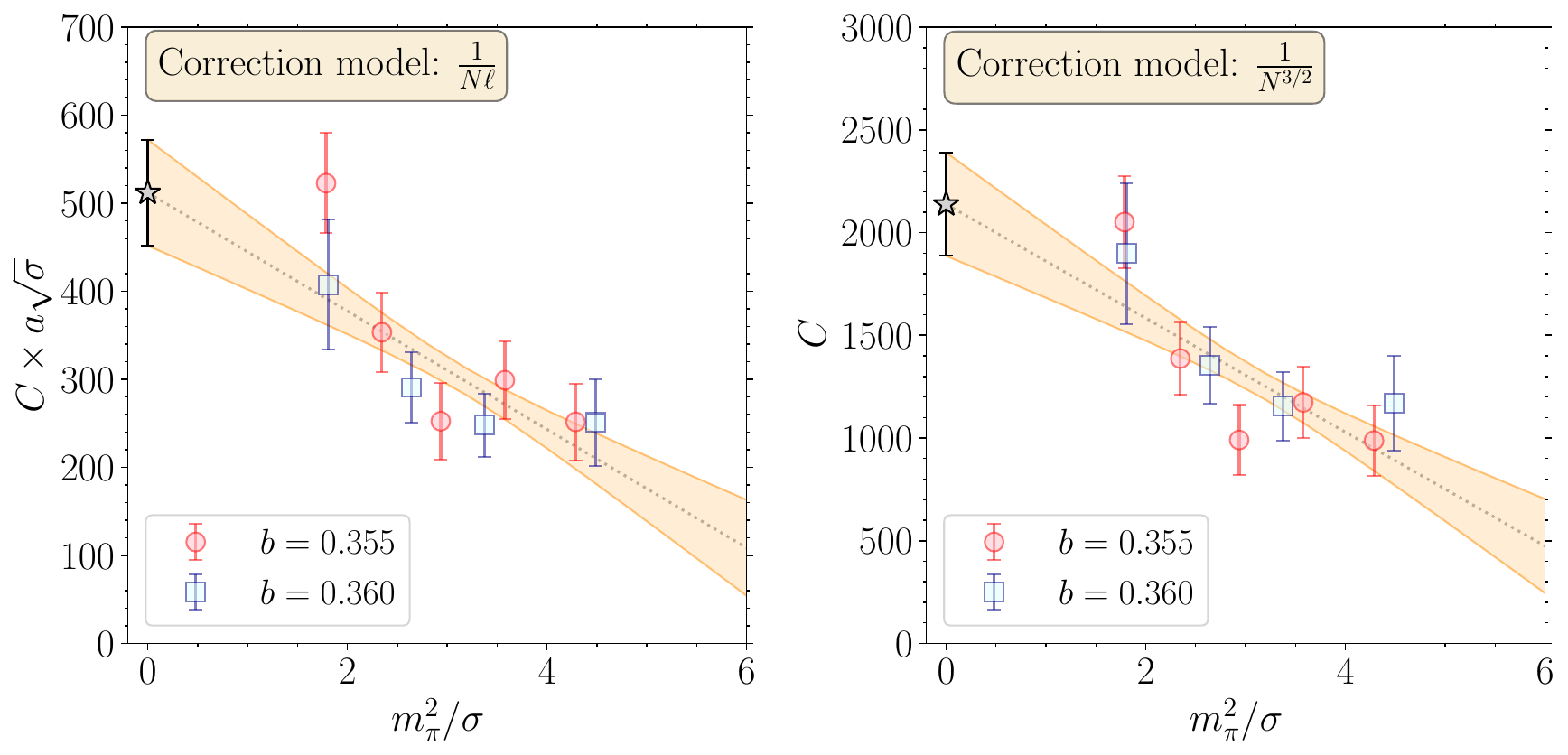}
\caption{Top panels: Large-$N$ extrapolations of the rescaled energy shift $\Delta \varepsilon$ assuming linear corrections in $1/N^{3/2}$ (inspired by L\"uscher formula). Obtained best fits have reduced chi-squared ranging from 0.48 to 1.37, corresponding to $p$-values ranging from 25\% to 70\%. Bottom panels: pion mass behavior of the pre-factor $C$ of the observed $1/N^{3/2}$ finite-$N$ dependence according to two different models. These best fits have a reduced chi-squared of, respectively, 1.59 (left) and 1.39 (right), corresponding to $p$-values of 13\% and 20\%.}
\label{fig:energy_shift_Ninf}
\end{figure}
 
The extraction of the scattering length from the energy shift proceeds via the L\"uscher formula.
As discussed in Sec.~\ref{sec:scatlength_Luscher}, at leading order:
\beq
\Delta \varepsilon = 4\pi \frac{Na_0}{m_\pi} \left[1+\mathcal{O}\left(\frac{1}{N\ell}\right)\right],
\eeq
i.e., the rescaled energy shift extracted from $\mathcal{R}_2(\tau)$ is equal to $4\pi N a_0 / m_\pi$ modulo corrections that vanish as $1/(N\ell)$. Based on this observation, as a first approximation, we assumed that the leading finite-$N$ corrections for TEK are dominated by those that correspond to the ones emerging from the expected finite volume corrections in the original theory, with the identification $L \rightarrow \sqrt{N}$. According to this assumption, our data should be fitted to an expression containing a linear correction in $1/N^{3/2}$, as follows:
\beq\label{eq:finite-N_effects_TEK}
\Delta \varepsilon = 4\pi \frac{Na_0}{m_\pi}\left(1+\frac{C}{N^{3/2}}\right) \, .
\eeq
Given that, for $b=0.355$ and $b=0.360$, the scattering length was computed for several values of $N$ for each $\kappa$, we performed individual fits to each set of $(b,\kappa)$ ensembles according to~\eqref{eq:finite-N_effects_TEK}. As it can be seen from the top panels of Fig.~\ref{fig:energy_shift_Ninf}, this ansatz describes our data very well, and allows a controlled extrapolation towards $N\to \infty$ in all cases. We also verified that perfectly consistent results are obtained leaving the exponent of $1/N$ as a free parameter, with best fit results for the free exponent always compatible with $3/2$.

Concerning the three finest lattice spacings, corresponding to $b=0.365,0.370,0.375$, we only have one value of $N=841$ at our disposal. Thus, in this case we estimated the size of finite effective volume effects from the $b=0.355$ and $b=0.360$ fits, and subtracted them from the $N=841$ data at $b=0.365,0.370,0.375$. The obtained results for $C$ exhibit a linear behavior as a function of $m_\pi^2/\sigma$ that points toward a finite chiral limit for this coefficient. Moreover, they seem to be affected by very mild lattice artifacts. This is sensible, as we are working with large enough volumes that allow for an effective separation between the ultra-violet cut-off $\Lambda_{\scriptscriptstyle{\rm UV}}=a^{-1}$ and the infra-red one $\Lambda_{\scriptscriptstyle{\rm IR}} = (aL)^{-1}=(a\sqrt{N})^{-1}=\Lambda_{\scriptscriptstyle{\rm UV}}/\sqrt{N}$. Thus, it is reasonable that finite-volume effects are to a large extent insensitive to lattice artifacts. Assuming them to be negligible, a joint fit of all $b=0.355$ and $b=0.360$ results for $C$ yields an excellent description of the data, see the bottom panels of Fig.~\ref{fig:energy_shift_Ninf}.

Due to the observed mild dependence on $a$, we cannot discriminate between a ``L\"uscher'' correction $\propto 1/(N\ell) = 1/(a N^{3/2})$ and a ``pure-$N$'' correction simply proportional to $1/N^{3/2}$ with no factors of $a$, cf.~the bottom panels of Fig.~\ref{fig:energy_shift_Ninf}. For this reason, we assumed both models to quantify and subtract finite-$N$ effects affecting the energy shift determinations obtained at $b=0.365,0.370,0.375$, where only one value of $N$ is available. The two $N=\infty$ estimates obtained from the two models are then combined using a standard AIC (Akaike Information Criterion) analysis~\cite{Jay:2020jkz}. More precisely, we subtracted the expected finite-volume effects computed from the two models from the $N=841$ data, and obtained the final result combining these two $N=\infty$ estimates via a weighted average. The weight of each model was computed as $\omega_i \propto\exp\left\{-\chi_i^2/2\right\}$ (with $\chi_i^2$ the total chi-squared of the fits depicted in the bottom panels of Fig.~\ref{fig:energy_shift_Ninf}) since both models have the same number of parameters and data points. To quote the final error, both statistical and systematic uncertainties were taken into account. The former were computed from the weighted average of the individual squared statistical uncertainties, while the latter were computed from the weighted average of the squared deviations between each $N=\infty$ determination and the AIC weighted average. In the end, the relative difference between the obtained $N=\infty$ results and the direct $N=841$ determinations at $b=0.365,0.370,0.375$ ranges from $\sim 4\%$ to $\sim 7\%$, thus signaling that the $N=841$ data are only affected by small residual finite volume effects. Final results for the large-$N$ scattering length $4\pi N a_0 / m_\pi$ are summarized in Tab.~\ref{tab:a0_final_res}.

\begin{table}[!t]
\small
\begin{center}
\begin{tabular}{|c|c|c|c|}
\hline
$b$ & $\kappa$ & $4\pi N a_0/(a^2 m_\pi)$ & $C$\\
\hline
0.355 & 0.1575 & 139.4(2.8) & 988(171)\\
0.355 & 0.1585 & 159.1(3.1) & 1173(173)\\
0.355 & 0.1592 & 181.2(3.2) & 990(170)\\
0.355 & 0.1600 & 202.6(3.1) & 1388(178)\\
0.355 & 0.1607 & 226.6(4.2) & 2052(223)\\
\hline
0.360 & 0.1560 & 185.9(4.3) & 1168(231)\\
0.360 & 0.1570 & 225.0(3.2) & 1155(167)\\
0.360 & 0.1577 & 258.7(3.4) & 1354(188)\\
0.360 & 0.1585 & 310.8(9.3) & 1898(343)\\
\cline{1-4}
\cline{1-3}
0.365 & 0.1567 & 428.3(15.7) & \multicolumn{1}{c}{} \\
\cline{1-3}
0.370 & 0.1543 & 413.9(14.1) & \multicolumn{1}{c}{} \\
\cline{1-3}
0.375 & 0.1527 & 415.9(20.9) & \multicolumn{1}{c}{} \\
\cline{1-3}
\end{tabular}
\end{center}
\caption{Summary of the final results for the infinite-volume large-$N$ $\pi^+\pi^+$ scattering length.}
\label{tab:a0_final_res}
\end{table}

\subsection{The pion-mass dependence of the scattering length}

\begin{figure}[!t]
\centering
\includegraphics[scale=0.57]{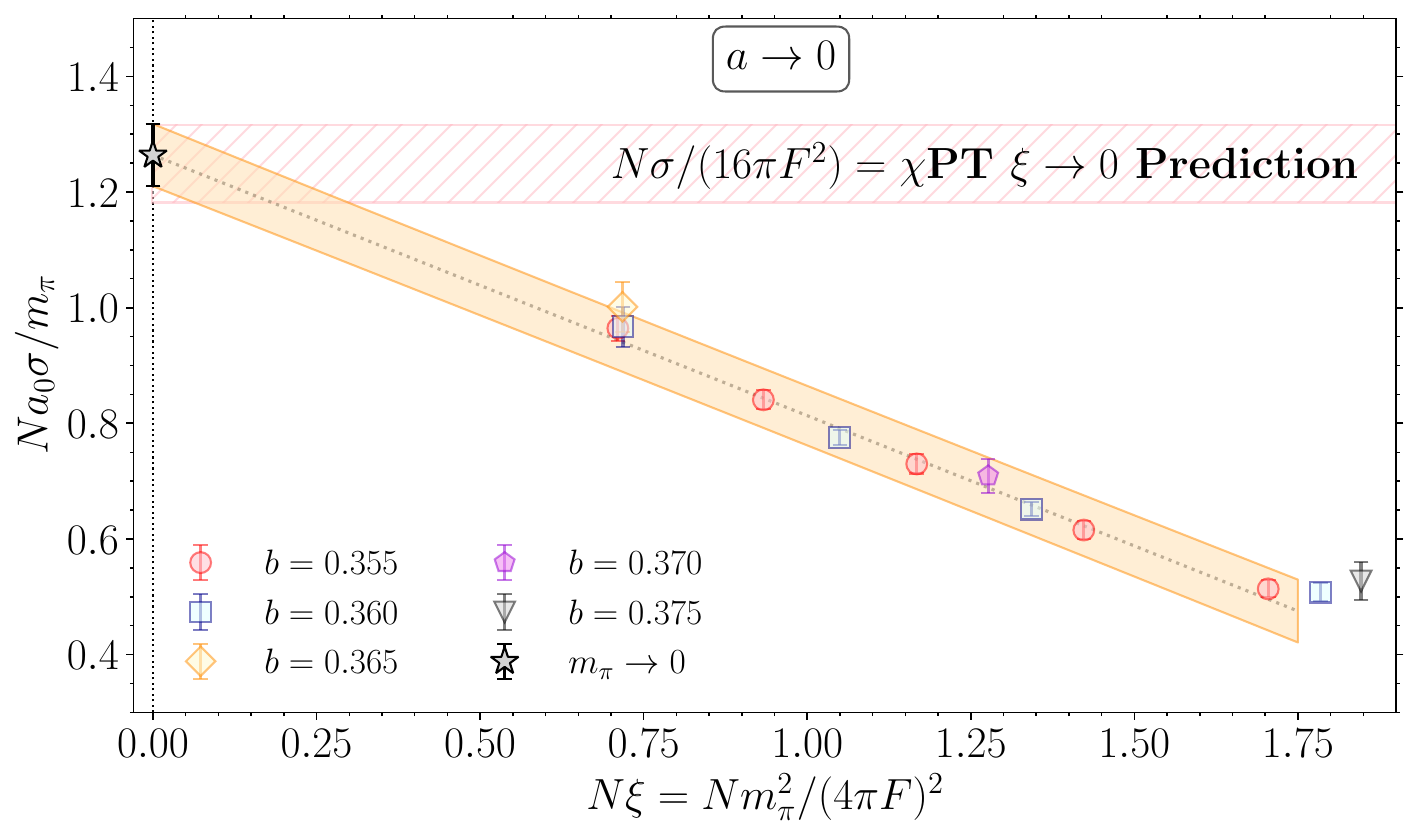}
\includegraphics[scale=0.57]{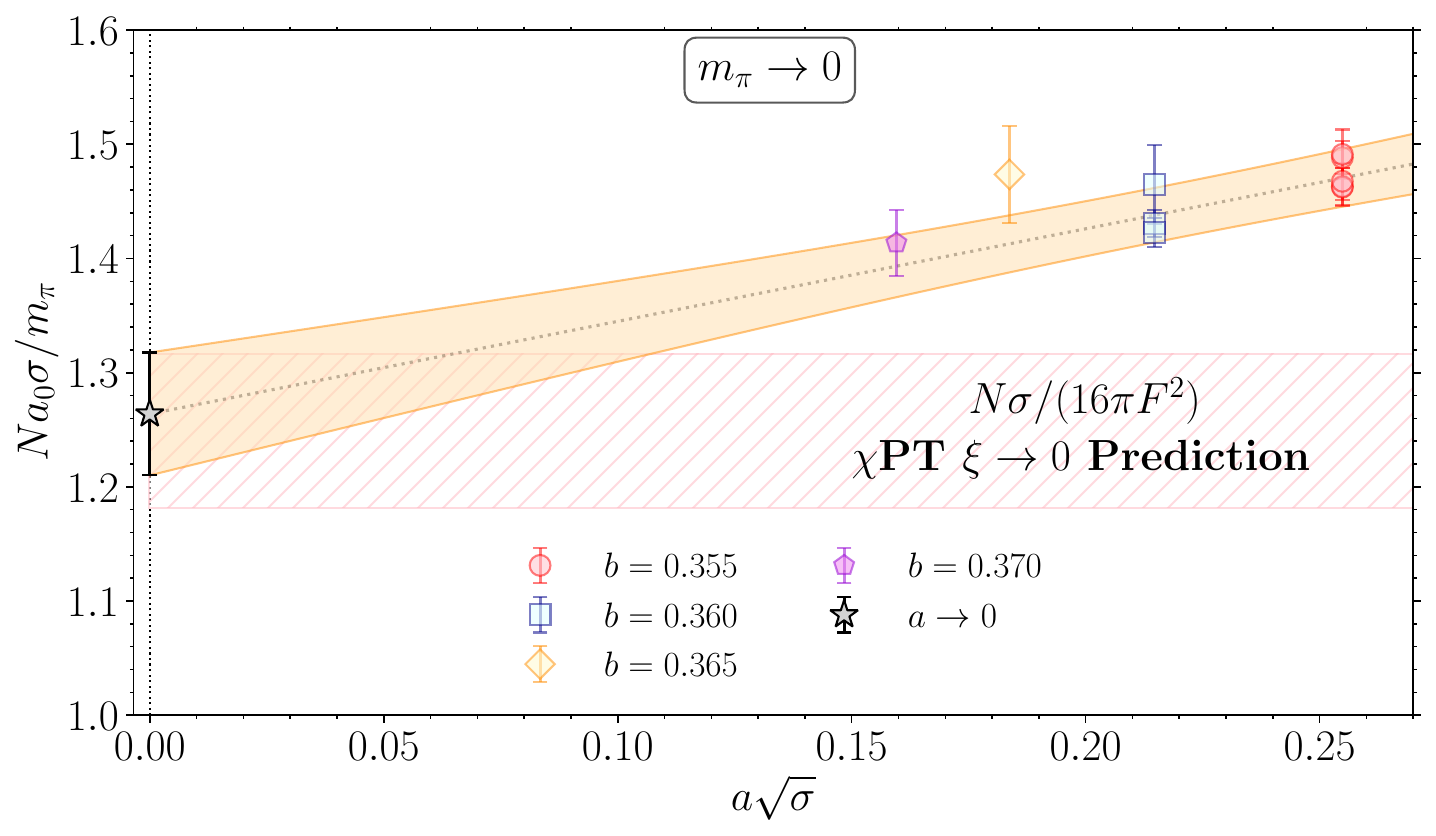}
\caption{Top panel: pion-mass dependence of the scattering length in the continuum limit. Bottom panel: lattice artifacts affecting the chiral limit of the scattering length. The shaded band represents in both cases the LO $\chiPT$ prediction for $N a_0\sigma/m_\pi$, expected to hold in the continuum and chiral limit. The chiral-continuum fit has a reduced chi-squared of 1.32 ($p$-value $\sim 23.5\%$).}
\label{fig:chircont_lim}
\end{figure}

We are now ready to study the pion-mass dependence of the scattering length, and compare it with the $\U(\Nf)$ $\chiPT$ prediction in Eq.~\eqref{eq:a0_chiPT_expansion}. Let us begin our analysis by considering the scattering length $Na_0/m_\pi$ in units of the string tension $\sigma$. We perform a global chiral-continuum best fit of our data for the scattering length assuming a fit function of the type:
\beq\label{eq:global_chircont_fit}
\frac{N a_0}{m_\pi}\sigma = A_0 \left(1 - A_1 N \xi + A_2 \, a\sqrt{\sigma}\right),
\eeq
where
\beq
N \xi = \left( \frac{m_\pi}{4 \pi F/\sqrt{N}} \right)^2 = \left[\frac{1}{4\pi} \times am_\pi \times \frac{\sqrt{8t_1}}{a} \times \frac{1}{\sqrt{8t_1\sigma}} \times \frac{\sqrt{N\sigma}}{F} \right]^2 .
\eeq
Here, $\sqrt{8t_1\sigma}$ is the value in Eq.~\eqref{eq:t1_sigma_conv}, while $F/\sqrt{N\sigma}$ stands for the continuum and chiral limit of the large-$N$ pion decay constant obtained from the TEK model in Ref.~\cite{Bonanno:2025hzr}:
\beq\label{eq:F_TEK_result}
\frac{1}{\sqrt{\sigma}} \frac{F}{\sqrt{N}} = 0.1262(34).
\eeq
In the right-hand side of Eq.~\eqref{eq:global_chircont_fit}, the first two terms are directly inspired from the expected pion-mass dependence predicted by NLO $\chiPT$, while the last contribution is an $\mathcal{O}(a)$ lattice artifact term introduced by the employed Wilson discretization. The results of this global fit are shown in Fig.~\ref{fig:chircont_lim}:
\begin{itemize}
\item The top panel shows the pion-mass dependence of the scattering length in the continuum limit, i.e., the continuum curve $A_0 ( 1 - A_1 N \xi)$ and the continuum-extrapolated points $\frac{N a_0}{m_\pi}\sigma - A_0A_2 \, a\sqrt{\sigma}$.
\item The bottom panel shows instead the lattice-spacing dependence of the chiral limit of the scattering length, i.e., the curve $A_0 (1 + A_2\, a \sqrt{\sigma})$ and the chiral-extrapolated points $\frac{N a_0}{m_\pi}\sigma + A_0A_1 N\xi$.
\end{itemize}

As it can be seen, both limits are under control: lattice artifacts in the chiral limit amount to a $\sim 15$\% effect at most (estimated as the relative difference between the values at the coarsest lattice spacing and in the continuum limit), and the chiral-continuum limit of $Na_0/m_\pi$ perfectly agrees with the expected LO $\chiPT$ prediction $N/(16 \pi F^2)$, computed using the TEK determination of $F/\sqrt{N}$ in Eq.~\eqref{eq:F_TEK_result}:
\beq
A_0 = 1.264(54), \qquad \qquad \text{LO $\chiPT$ prediction} = \frac{N\sigma}{16\pi F^2} = 1.249(67).
\eeq
The slope term $-A_1 N \xi$ thus allows to determine the combination of LECs:
\beq
A_1 = \frac{\Lpi-\bar{\ell}_3/2}{N} = 0.357(18).
\eeq
In order to extract $\Lpi$, one needs a determination of $\bar{\ell}_3$, parameterizing the sub-leading quark-mass dependence of the pion mass at large $N$. In the previous TEK study~\cite{Bonanno:2025hzr}, $m_\pi^2$ is observed to be perfectly described by a linear behavior in the whole explored range of quark masses, thus pointing out that $\bar{\ell}_3$ is much smaller than our current error bars. This is consistent with previous large-$N$ lattice~\cite{Hernandez:2019qed,DeGrand:2023hzz} and model~\cite{Pich:2002xy} estimations of this LEC, pointing out that it is very small. In particular, expressing $\bar{\ell}_3/N = -256\pi^2 (2L_8 - L_5)/N$ in terms of the 3-flavor LECs $L_5$ and $L_8$, one finds that $\bar{\ell}_3$ is small due to an accidental cancellation between $L_5$ and $L_8$, since $2L_8 \approx L_5$. In any case, we can use the data of~\cite{Bonanno:2025hzr} (combined with the updated $m_\pi$ values provided here) to obtain a bound on this LEC. Our large-$N$ TEK estimation of $\bar{\ell}_3/N$ is detailed in Appendix~\ref{app:l3}, and we obtain the following bound: $\bar{\ell}_3/N=-0.022(22)$. This determination finally allows to isolate $\Lpi/N$.

\begin{figure}[!t]
\centering
\includegraphics[scale=0.55]{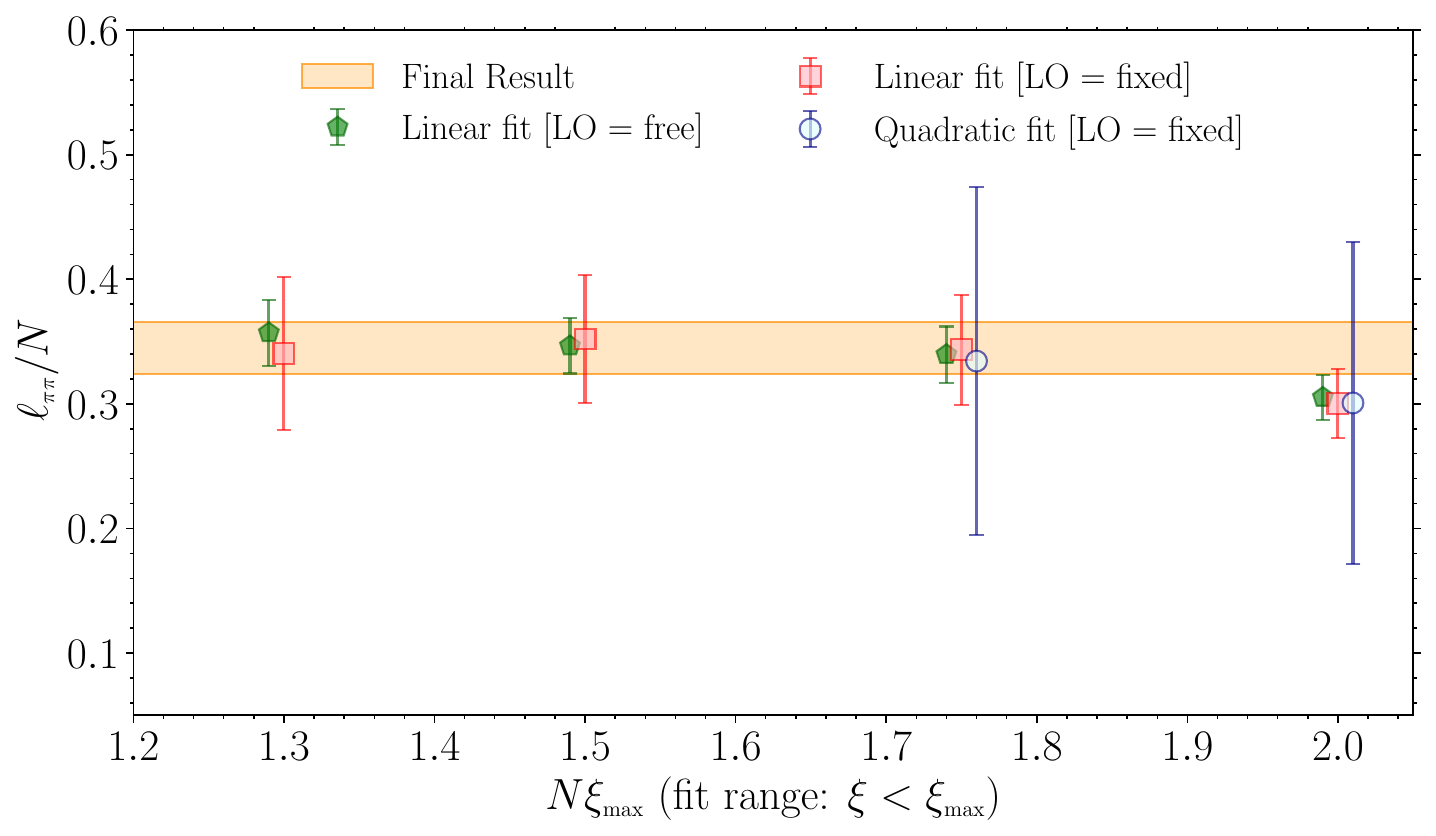}
\caption{Study of possible systematic effects affecting our determination of $\Lpi/N$.}
\label{fig:syst_lpipi}
\end{figure}

To provide our final result for $\Lpi$, we performed several fits to assess the magnitude of possible systematic uncertainties. In particular, we tried several fit ranges with different upper bounds, and we also tried to fit the quantity $F^2 a_0 / m_\pi$ fixing its chiral limit to $1/(16\pi)$ (as predicted by $\chiPT$). In the latter case, for the two widest fit ranges we also tried to add a quadratic term in $N\xi$ to the global fit function. The outcome of these different best fits is summarized in Fig.~\ref{fig:syst_lpipi}. As it can be seen, $\Lpi/N$ turns out to be pretty stable against the different fit strategies employed, pointing out that our estimate is robust. In the end, we quote the following final results, represented in Fig.~\ref{fig:syst_lpipi} as a shaded band:
\beq
\frac{\Lpi}{N} = 0.345(21).
\eeq
Let us conclude by commenting that, using directly the string tension to set the scale (without passing through the gradient flow scale $t_1$) would lead to the perfectly compatible results: $A_0=1.310(82)$ and $A_1 = 0.300(27)$ (differing by about 1 standard deviation with the ones obtained via $t_1$).

\subsection{The \texorpdfstring{$1/N$}{1/N} expansion of \texorpdfstring{$\Lpi$}{lpipi}}
The goal of this section is to combine the large-$N$ TEK result for $\Lpi/N$ obtained in this study with the finite-$N$ results obtained in the $\Nf=4$ $\pi^+\pi^+$ scattering lattice study of Ref.~\cite{Baeza-Ballesteros:2022azb} to constrain the coefficients of the $1/N$ expansion of this LEC.

\begin{figure}[!t]
\centering
\includegraphics[scale=0.55]{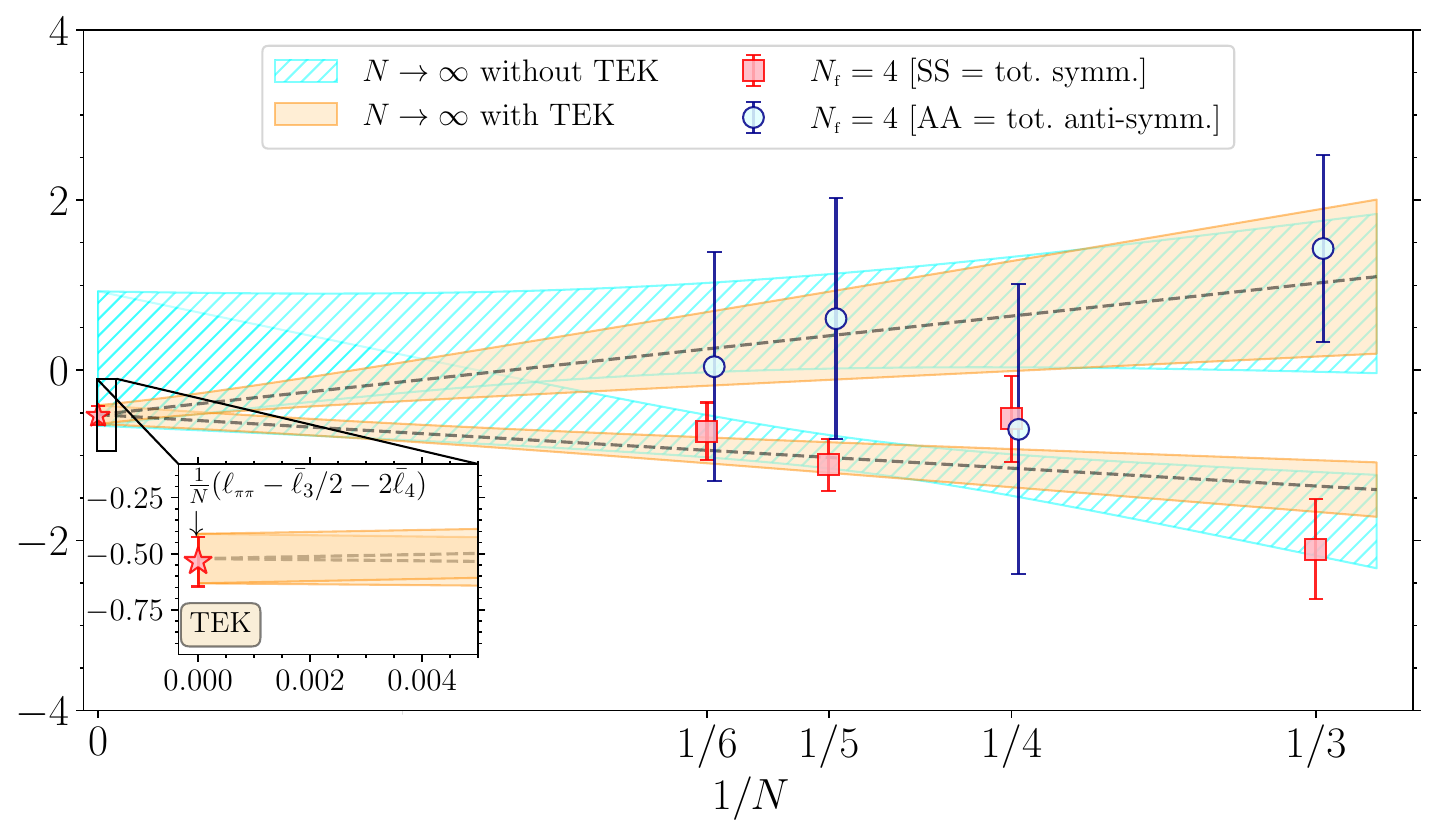}
\caption{Calculation of the $1/N$ expansion of the combination of LECs given by $\frac{\Lpi - \bar{\ell}_3/2 - 2 \bar{\ell}_4}{N}$, obtained combining the large-$N$ TEK determination with the finite-$N$ results of~\cite{Baeza-Ballesteros:2022azb}.}
\label{fig:Ndep_LEC}
\end{figure}

In that study, the authors computed the equivalent of $\Lpi$ in two different channels, the totally-symmetric channel SS (the $\Nf=4$ analogue of the isospin-2 channel considered in this study), and the totally-antisymmetric channel AA (only existing for $\Nf\ge 4$). The LECs of these two channels, $\lSS\sim\mathcal{O}(N)$ and $\lAA\sim\mathcal{O}(N)$ possess the same large-$N$ limit, and differ only for sub-leading terms in the $1/N$ expansion:\footnote{Note that they are related to the LECs $L_{\sstrm{SS}}$ and $L_{\sstrm{AA}}$ discussed in Ref.~\cite{Baeza-Ballesteros:2022azb} via $\lSS=(16 \pi)^2 L_{\sstrm{SS}}$ and $\lAA=(16 \pi)^2 L_{\sstrm{AA}}$.}
\beq
\label{eq:lSS_def}
\frac{\lSS}{N} &=& \ell_{\sst{\infty}} + \frac{1}{N}(\Nf \ell^{\sst{(1)}}_{\sstrm{c}} + \ell^{\sst{(1)}}_{\sstrm{a}}) + \mathcal{O}\left(\frac{1}{N^2}\right),\\
\label{eq:lAA_def}
\frac{\lAA}{N} &=& \ell_{\sst{\infty}} + \frac{1}{N}(\Nf \ell^{\sst{(1)}}_{\sstrm{c}} - \ell^{\sst{(1)}}_{\sstrm{a}}) + \mathcal{O}\left(\frac{1}{N^2}\right).
\eeq
Since the authors of~\cite{Baeza-Ballesteros:2022azb} employed the alternative variable $x=m^2_\pi/(16\pi^2 F_\pi^2)$ in the $\chi$PT expansion, the quantity $\lSS$ parameterizing the NLO term in the chiral expansion of $a_0/m_\pi$ in the SS channel is equal to:
\beq
\lSS = \Lpi - \frac{1}{2}\bar{\ell}_3 - 2 \bar{\ell}_4,
\eeq
cf.~also Eqs.~(90) and (92) in Ref.~\cite{FlavourLatticeAveragingGroupFLAG:2021npn}. Given that the large-$N$ limit of $\bar{\ell}_4/N=0.446(55)$ has been determined from the TEK model in Ref.~\cite{Bonanno:2025hzr}, we are able to determine:
\beq
\frac{\Lpi - \bar{\ell}_3/2 - 2 \bar{\ell}_4}{N}\bigg\vert_{\rm TEK} = \ell_{\sst{\infty}},
\eeq
and use this value, along with the finite-$N$ data for $\lSS$ and $\lAA$ of~\cite{Baeza-Ballesteros:2022azb}, to determine $\ell^{\sst{(1)}}_{\sstrm{c}}$ and $ \ell^{\sst{(1)}}_{\sstrm{a}}$ with a global fit. Thanks to the anti-correlation between the $\Nf$-dependent and the $\Nf$-independent terms appearing in the sub-leading contribution to $\lSS$ and $\lAA$, cf.~Eqs.~\eqref{eq:lSS_def} and~\eqref{eq:lAA_def}, we will also be able to completely characterize the $\Nf$-dependence of the $\mathcal{O}(1/N)$ term of the $1/N$ expansion of $\Lpi$.

\begin{table}[!t]
\small
\begin{center}
\begin{tabular}{|c|c|c|}
\hline
$N$ & $\lSS/N$ & $\lAA/N$ \\
\hline
3 & -2.11(59) & 1.4(1.1)    \\
4 & -0.57(51) & -0.69(1.71) \\
5 & -1.11(30) & 0.61(1.41)  \\
6 & -0.72(34) & 0.04(1.35)  \\
\hline
$\infty$ & \multicolumn{2}{c|}{-0.54(11)}\\
\hline
\end{tabular}
\end{center}
\caption{Finite-$N$ $\Nf=4$ determinations of $\lSS/N$ and $\lAA/N$ of Ref.~\cite{Baeza-Ballesteros:2022azb}, along with the common $N=\infty$ TEK result obtained in this study.}
\label{tab:LSS_LAA_vs_N}
\end{table}

This calculation is shown in Fig.~\ref{fig:Ndep_LEC}, where it is compared with the best fit performed in~\cite{Baeza-Ballesteros:2022azb} without the TEK determination. The fitted data can be found in Tab.~\ref{tab:LSS_LAA_vs_N}. The improvement in accuracy thanks to the inclusion of the new TEK $N=\infty$ determination is dramatic, and amounts to more than an order of magnitude for the large-$N$ limit $\ell_{\sst{\infty}}$. The final results read:
\beq
\ell_{\sst{\infty}}=-0.52(11), \qquad \qquad \ell^{\sst{(1)}}_{\sstrm{c}} =0.26(36) , \qquad \qquad \ell^{\sst{(1)}}_{\sstrm{a}}=-3.58(1.37).
\eeq
Combining these results with the $\U(\Nf)$ $\chiPT$ $1/N$ expansion of $\bar{\ell}_3/N$ reported in Eq.~\eqref{eq:l3_U_Nf} in Appendix~\ref{app:l3} and of $\bar{\ell}_4/N = 0.446(55) - 0.44(15)\Nf/N$ reported in Ref.~\cite{Bonanno:2025hzr}, we finally find:
\beq
\frac{\Lpi}{N} = 0.345(21) - \frac{1}{N}\left[\Nf \, 1.43(50) + 3.6(1.4) \right] + \mathcal{O}\left(\frac{1}{N^2}\right).
\eeq

\subsection{Comparison with bootstrap bounds}

In Refs.~\cite{Albert:2023seb,Albert:2026xyz}, bootstrap bounds for the couplings $\bar{\ell}_1/N$ and $\bar{\ell}_2/N$ have been derived in large-$N$ QCD in the chiral limit. Our determination of $\frac{\Lpi}{N} = \frac{4}{3N}\left(\bar{\ell}_1 + 2\bar{\ell}_2\right)$ provides a further constraint on the possible values that these two LECs can take in this theory. It is thus interesting to compare it with the results of Refs.~\cite{Albert:2023seb,Albert:2026xyz}. To do so, we first need to convert our result in the same notation adopted in~\cite{Albert:2023seb,Albert:2026xyz}.\footnote{We are grateful to J.~Albert and L.~Rastelli for correspondence on this point.} In particular, let us introduce the following quantities:
\beq
\gtwo = \frac{1}{48\pi^2}\left(\bar{\ell}_1+3\bar{\ell}_2\right)\left[\frac{m_\rho^{\chir}}{F}\right]^2, \qquad \qquad \gtwop = \frac{1}{12\pi^2}\bar{\ell}_2 \left[\frac{m_\rho^{\chir}}{F}\right]^2 ,
\eeq
where $m_\rho^{\chir}$ is the large-$N$ $\rho$-meson mass in the chiral limit. Using the TEK determination $m_\rho^{\chir}/\sqrt{\sigma}=1.783(47)$ obtained in the continuum limit in Ref.~\cite{Bonanno:2025hzr}, and the result for $F/\sqrt{N\sigma}$ in Eq.~\eqref{eq:F_TEK_result} (again, obtained from the TEK model in the continuum limit in~\cite{Bonanno:2025hzr}), one obtains:
\beq\label{eq:TEK_constraint}
4 \gtwo - \gtwop = \frac{\Lpi}{16\pi^2}\left[\frac{m_\rho^{\chir}}{F}\right]^2 = 0.436(42).
\eeq

\begin{figure}[!t]
\centering
\includegraphics[scale=0.55]{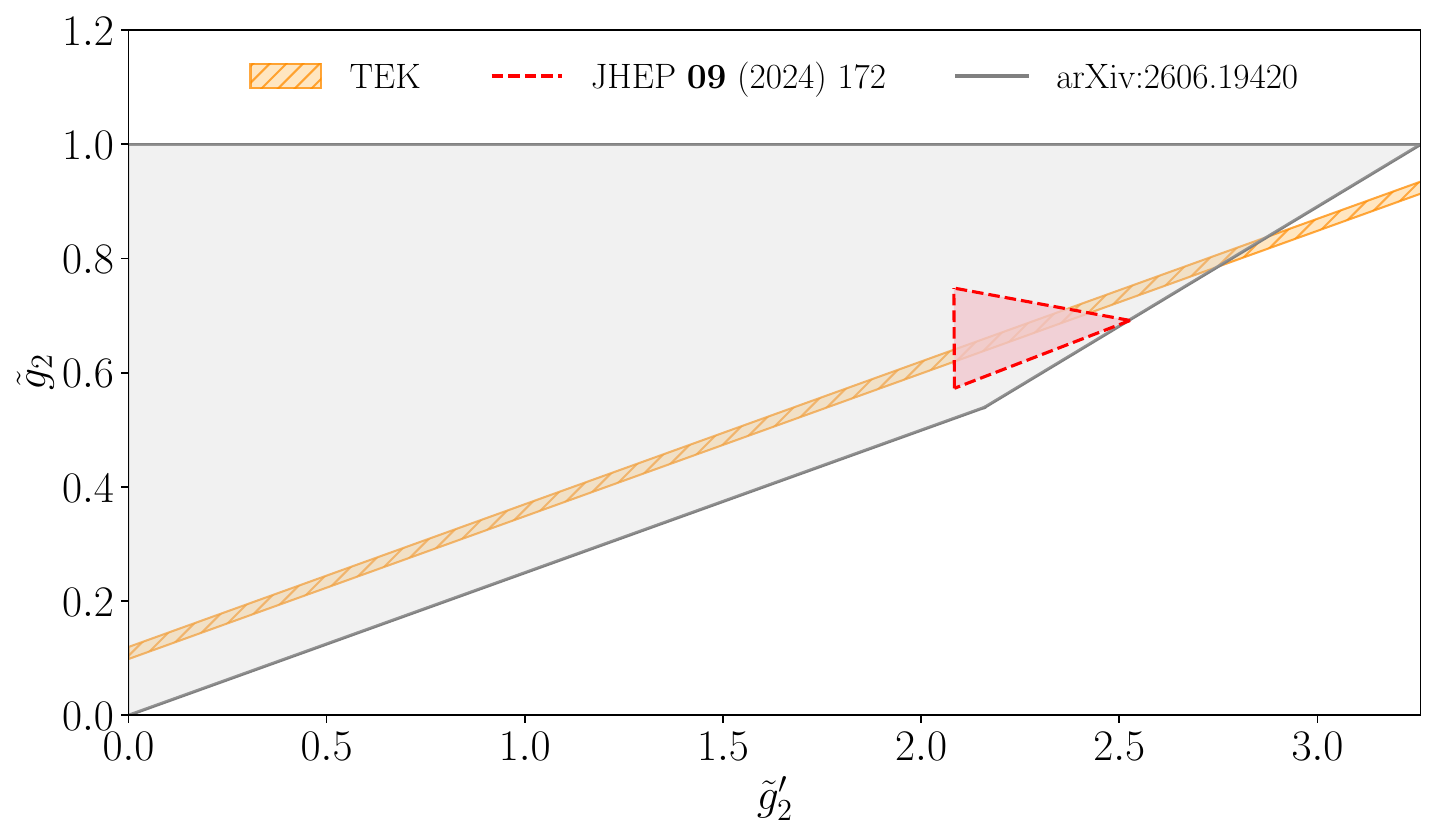}
\caption{Bounds on $\gtwo$ and $\gtwop$ obtained in~\cite{Albert:2023seb,Albert:2026xyz}, compared with the constraint provided by the determination of $\Lpi/N$ obtained in this study, cf.~Eq.~\eqref{eq:TEK_constraint}.}
\label{fig:comp_bootstrap}
\end{figure}

The summary of the bounds on $\gtwo$ and $\gtwop$ of~\cite{Albert:2023seb,Albert:2026xyz} and of the constraint in Eq.~\eqref{eq:TEK_constraint} on $\gtwo$ and $\gtwop$ provided by our large-$N$ determination of $\Lpi/N$ is shown in Fig.~\ref{fig:comp_bootstrap}. Our result for $\Lpi/N$ translates into a linear strip in the $\gtwo-\gtwop$ plane. The bounds derived in~\cite{Albert:2026xyz} were obtained considering three pion form factors --- the vector-current two-point function, the pion vector form factor, and the pion scattering amplitude --- and are based both on general theoretical properties of these correlators (analyticity, unitarity, asymptotic scaling) and on Shifman--Vainshtein--Zakharov-like sum rules to bound the pion charge radius and the scale at which asymptotic freedom sets in. Note that a key ingredient to the latter procedure was the usage of the lattice results for the pion decay constant and the $\rho$ mass in the large-$N$ and chiral limits obtained from the TEK study~\cite{Bonanno:2025hzr}. The triangle, derived in~\cite{Albert:2023seb}, is instead a bootstrap
bound derived from the masses and couplings of the first two mesons on the leading Regge trajectory (i.e., the $\rho$ and the $f_2$) fixed to their real-world values.\footnote{We note that the large-$N$ $\rho$ mass in the chiral limit $m_{\rho}^{\chir}/\sqrt{\sigma}=1.783(47)$ is very close to the experimental value, $m_\rho^{\scriptscriptstyle{\rm (exp)}}/\sqrt{\sigma}\vert_{\rm phys}=1.742(35)$. The latter value is obtained using $\sqrt{\sigma}\vert_{\rm phys}=445(9)$ MeV from~\cite{Bulava:2024jpj} (determined via 2+1 lattice QCD simulations, and extrapolated toward the physical point), and $m_\rho^{\scriptscriptstyle{\rm (exp)}}=775.26(25)$ MeV from the PDG~\cite{ParticleDataGroup:2024cfk}.} As it can be observed, there is very good compatibility among all of these bounds. This is a quite rewarding achievement given the very different frameworks and assumptions adopted for their derivation.

\section{Conclusions}\label{sec:conclu}

In this paper we have presented the TEK calculation of the large-$N$ $\pi^+ \pi^+$ scattering length, which represents the first calculation of a scattering quantity in the large-$N$ limit with this framework. The scattering length has been computed using the L\"uscher finite-volume formalism, and the finite-volume energy shift of the 2-pion system has been extracted from the calculation of a suitable 4-point function.

The usage of several lattice parameters spanning several lattice spacings, quark masses and gauge-group ranks allowed us to take controlled continuum, chiral and large-$N$ limits. In the end, we are able to determine the chiral expansion of the quantity $Na_0/m_\pi$ up to $\mathcal{O}(m_\pi^4)$ corrections in the continuum limit:
\beq
N \frac{a_0}{m_\pi}\sigma = 1.264(54) \left[1 - 0.357(18) \, N\xi + \mathcal{O}(N^2\xi^2)\right], \qquad \qquad \xi = \frac{m_\pi^2}{(4\pi F)^2}.
\eeq
The chiral limit of $N \sigma a_0/m_\pi$ agrees with the LO $\chiPT$ prediction $N\sigma/(16\pi F^2) = 1.249(67)$ (evaluated using the TEK determination of $F/\sqrt{N}$ in units of the string tension). Moreover, the matching of the NLO term to $\chiPT$ allows the extraction of a new large-$N$ low-energy constant:
\beq
\frac{\Lpi}{N} = \frac{4}{3N}\left(\bar{\ell}_1 + 2\bar{\ell}_2\right) = 0.345(21),
\eeq
which turns out to be in very good agreement with recent S-matrix bootstrap bounds on the couplings $\gtwo$ and $\gtwop$, a combination of $\bar{\ell}_1$, $\bar{\ell}_2$, $m_\rho^\chir$ and $F$. Currently, our $\sim 10\%$ error on $(16\pi^2)(4\gtwo - \gtwop) = \Lpi (m_\rho^\chir/F)^2$ is determined by the $\sim 6\%$ uncertainty on $\Lpi$, and  the $\sim 5\%$ errors on $F^2$ and $\left(m_\rho^{\chir}\right)^2$. Since lattice artifacts for these quantities are observed to be reasonably mild, reducing the error on this combination further would require simulations closer to the chiral limit. This is numerically challenging, as it in turn requires very large values of $N$ (especially on finer lattice spacings), as both $\ell\sqrt{\sigma}=\sqrt{N} a \sqrt{\sigma}$ and $m_\pi \ell = a m_\pi\sqrt{N}$ have to be kept sufficiently large to avoid significant finite-size effects. In this respect, the main bottleneck is given by the inversion of the Dirac operator to compute fermion correlators, which is much more expensive than the generation of pure-gauge configurations.

Combining TEK results with previous finite-$N$ lattice QCD data for the pion-pion scattering length, we are able to constrain sub-leading corrections to $\Lpi/N$:
\beq
\frac{\Lpi}{N} = 0.345(21) - \frac{1}{N}\left[\Nf \, 1.43(50) + 3.6(1.4) \right] + \mathcal{O}\left(\frac{1}{N^2}\right).
\eeq
These results, along with the result obtained in the previous TEK study~\cite{Bonanno:2025hzr}:
\beq
\frac{F}{\sqrt{N\sigma}} = 0.1262(34)\left[1 - 0.46(7) \frac{\Nf}{N} + \mathcal{O}\left(\frac{1}{N^2}\right)\right] \, ,
\eeq
constitute a first-principles determination of the large-$N$ $\pi^+\pi^+$ scattering length $Na_0/m_\pi$ up to NLO both in the chiral and in the $1/N$ expansions.

This investigation opens several new research directions which could be pursued in the near future. The calculation of the pion-pion scattering length in the isospin-0 channel is an obvious next step, and would be extremely interesting for its relation with the fate of the $\sigma$ resonance in the large-$N$ limit. The study of other pion form factors such as those considered in Refs.~\cite{DeGrand:2024frm,Albert:2026xyz} would be extremely interesting too. Another intriguing development would be the study of $\rho\rho$ and $\pi\rho$ scattering. Apart from the extraction of the scattering length from 4-point functions using the L\"uscher finite-volume formalism, in these cases it would also be interesting to study suitable 3-point functions from which the trilinear couplings $g_{\pi\pi\rho}$, $g_{\pi\rho\rho}$ or $g_{\rho\rho\rho}$ can be extracted. These predictions would provide valuable inputs for complementary non-perturbative studies of large-$N$ QCD currently ongoing using bootstrap, dispersive and effective theory approaches.

\section*{Acknowledgements}
It is a pleasure to thank J.~Albert, M.~Bruno, M.~Cè, L.~Giusti, M.~Hansen, S.~Kim, M.~Mai, L.~Rastelli, S.~Sharpe, R.~Venugopalan, A.~Vichi, and R.~Zwicky for useful discussions. CB thanks the Albert Einstein Center of the University of Bern for supporting a visit during which this work was initiated. This work is partially supported by the Spanish Research Agency (Agencia Estatal de Investigaci\'on) through the grant IFT Centro de Excelencia Severo Ochoa CEX2020-001007-S and, partially, by the grants PID2021-127526NB-I00 and PID2024-160152NB-I00, both funded by MCIN/AEI/10.13039/501100011033. It is also partially funded by the European Commission – NextGenerationEU, through Momentum CSIC Programme: Develop Your Digital Talent. The work of FRL was supported in part by the Swiss National Science Foundation (SNSF) through grant No. 200021-236432, and the Platform for Advanced Scientific Computing (PASC) project “ALPENGLUE”. Numerical calculations have been performed on the \texttt{Finisterrae~III} cluster at CESGA (Centro de Supercomputaci\'on de Galicia), and on the Hydra and Ciclope clusters at IFT. We acknowledge HPC support by Emilio Ambite, staff hired under the Generation D initiative, promoted by Red.es, an organization attached to the Ministry for Digital Transformation and the Civil Service, for the attraction and retention of talent through grants and training contracts, financed by the Recovery, Transformation and Resilience Plan through the European Union (EU)’s Next Generation funds. We have also used the computational resource of the Miyabi-G system at the Joint Center for Advanced High Performance Computing (JCAHPC), which is jointly operated by the University of Tokyo and the University of Tsukuba.

\appendix
\section*{Appendix}
\section{Raw data}\label{app:raw_data}

The complete collection of raw data for the energy shift is provided in Tab.~\ref{tab:raw_data}.

\begin{table}[!t]
\begin{center}
\begin{tabular}{|c|c|c|c|}
\hline
$N$ & $b$ & $\kappa$ & $\Delta \varepsilon/a^2$ \\
\hline
289  &  0.355  &  0.1575  &  169.2(2.8) \\
361  &  0.355  &  0.1575  &  155.5(2.9) \\
529  &  0.355  &  0.1575  &  151.7(2.7) \\
841  &  0.355  &  0.1575  &  147.7(3.5) \\
1156  &  0.355  &  0.1575  &  140.1(4.1) \\
\hline
289  &  0.355  &  0.1585  &  197.0(3.2) \\
361  &  0.355  &  0.1585  &  184.8(3.2) \\
529  &  0.355  &  0.1585  &  176.7(2.7) \\
841  &  0.355  &  0.1585  &  165.8(3.7) \\
1156  &  0.355  &  0.1585  &  160.3(5.4) \\
\hline
289  &  0.355  &  0.1592  &  224.8(5.1) \\
361  &  0.355  &  0.1592  &  203.5(2.8) \\
529  &  0.355  &  0.1592  &  197.3(2.9) \\
841  &  0.355  &  0.1592  &  188.8(3.0) \\
1156  &  0.355  &  0.1592  &  186.4(5.9) \\
\hline
289  &  0.355  &  0.1600  &  263.9(6.0) \\
361  &  0.355  &  0.1600  &  238.7(4.8) \\
529  &  0.355  &  0.1600  &  227.0(3.5) \\
841  &  0.355  &  0.1600  &  213.9(3.3) \\
1156  &  0.355  &  0.1600  &  210.0(4.0) \\
\hline
289  &  0.355  &  0.1607  &  327.2(8.2) \\
361  &  0.355  &  0.1607  &  288.4(6.0) \\
529  &  0.355  &  0.1607  &  265.3(5.2) \\
841  &  0.355  &  0.1607  &  248.8(4.1) \\
1156  &  0.355  &  0.1607  &  234.2(5.6) \\
\hline
\end{tabular}
\begin{tabular}{|c|c|c|c|}
\hline
$N$ & $b$ & $\kappa$ & $\Delta \varepsilon/a^2$ \\
\hline
289  &  0.360  &  0.1560  &  231.7(5.5) \\
361  &  0.360  &  0.1560  &  217.9(4.5) \\
529  &  0.360  &  0.1560  &  202.2(2.5) \\
841  &  0.360  &  0.1560  &  198.7(4.6) \\
\hline
289  &  0.360  &  0.1570  &  279.5(6.3) \\
361  &  0.360  &  0.1570  &  265.2(4.5) \\
529  &  0.360  &  0.1570  &  244.1(2.2) \\
841  &  0.360  &  0.1570  &  237.5(2.5) \\
\hline
289  &  0.360  &  0.1577  &  332.3(9.3) \\
361  &  0.360  &  0.1577  &  312.8(6.8) \\
529  &  0.360  &  0.1577  &  284.7(3.3) \\
841  &  0.360  &  0.1577  &  274.0(2.3) \\
\hline
289  &  0.360  &  0.1585  &  426.3(12.6) \\
361  &  0.360  &  0.1585  &  412.9(14.8) \\
529  &  0.360  &  0.1585  &  357.5(5.1) \\
841  &  0.360  &  0.1585  &  337.6(9.7) \\
\hline
\hline
841  &  0.365  &  0.1567  &  462.5(11.7) \\
\hline
\hline
841  &  0.370  &  0.1543  &  440.2(9.3)  \\
\hline
\hline
841  &  0.375  &  0.1527  &  433.2(16.8) \\
\hline
\end{tabular}
\end{center}
\caption{Collection of raw data for the rescaled energy shift $\Delta \varepsilon$ in lattice units.}
\label{tab:raw_data}
\end{table}

\section{Energy shift extraction: additional plots}\label{app:additional_plot}

\begin{figure}[!t]
\centering
\includegraphics[scale=0.475]{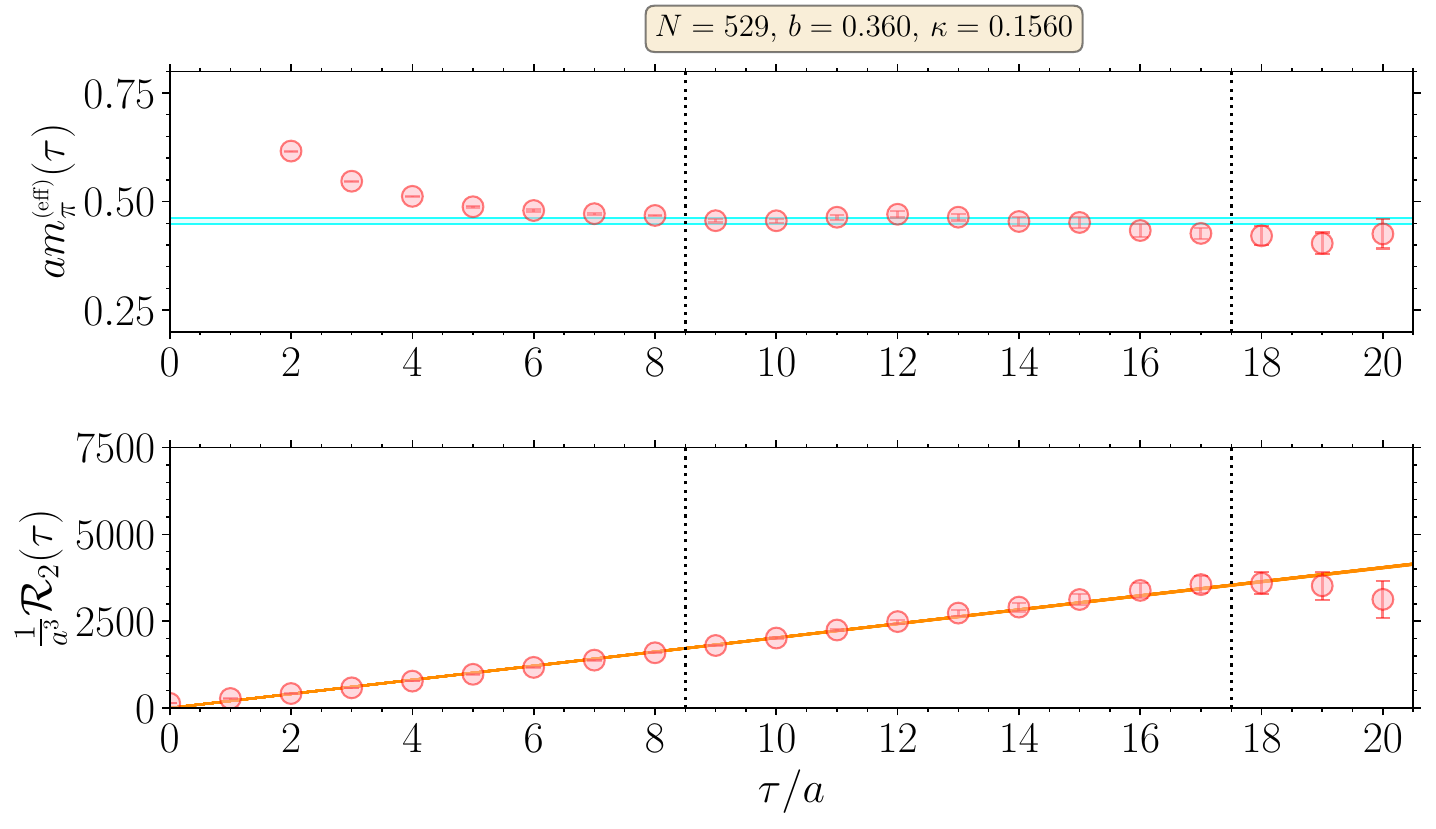}
\includegraphics[scale=0.475]{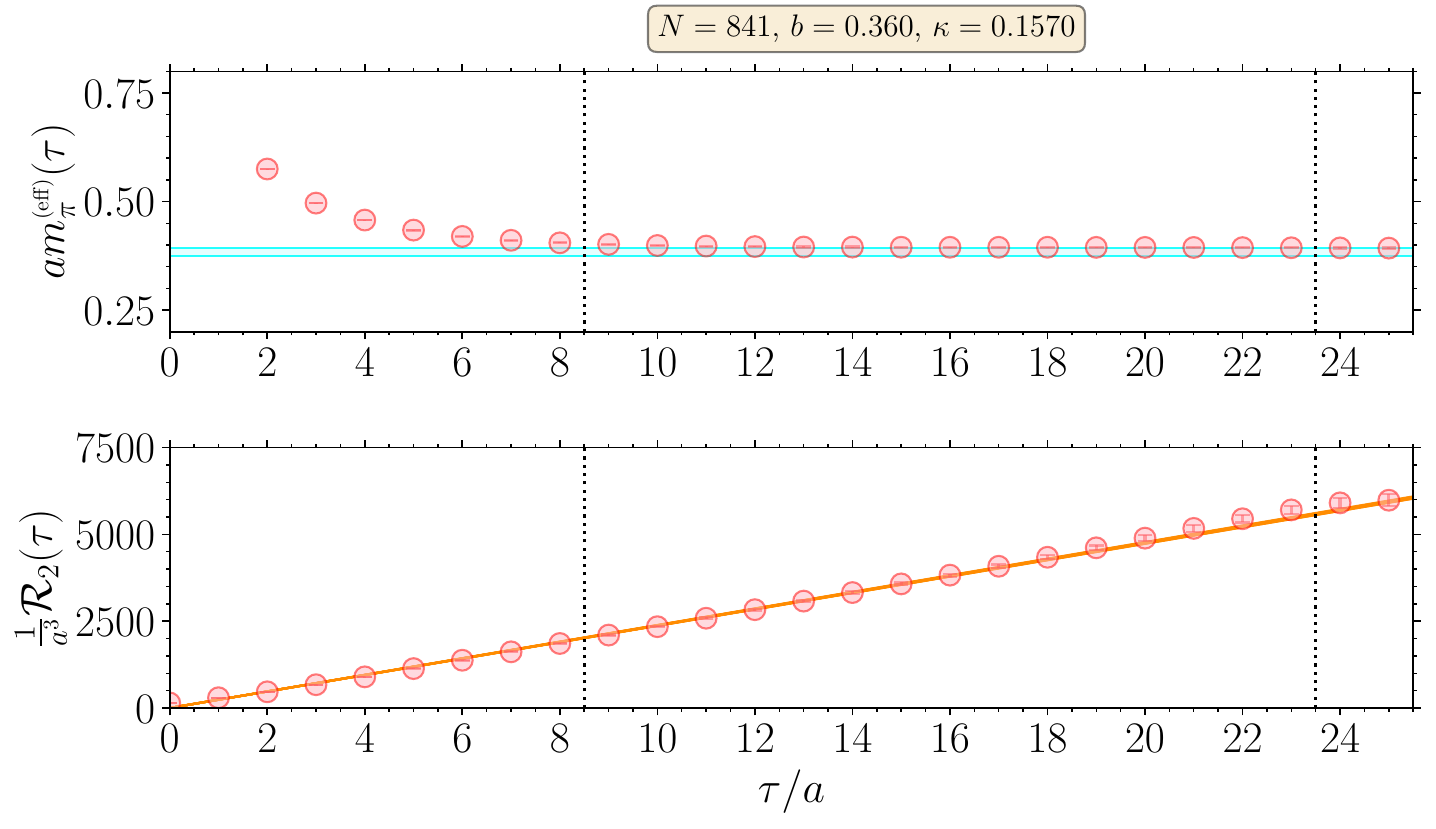}
\includegraphics[scale=0.475]{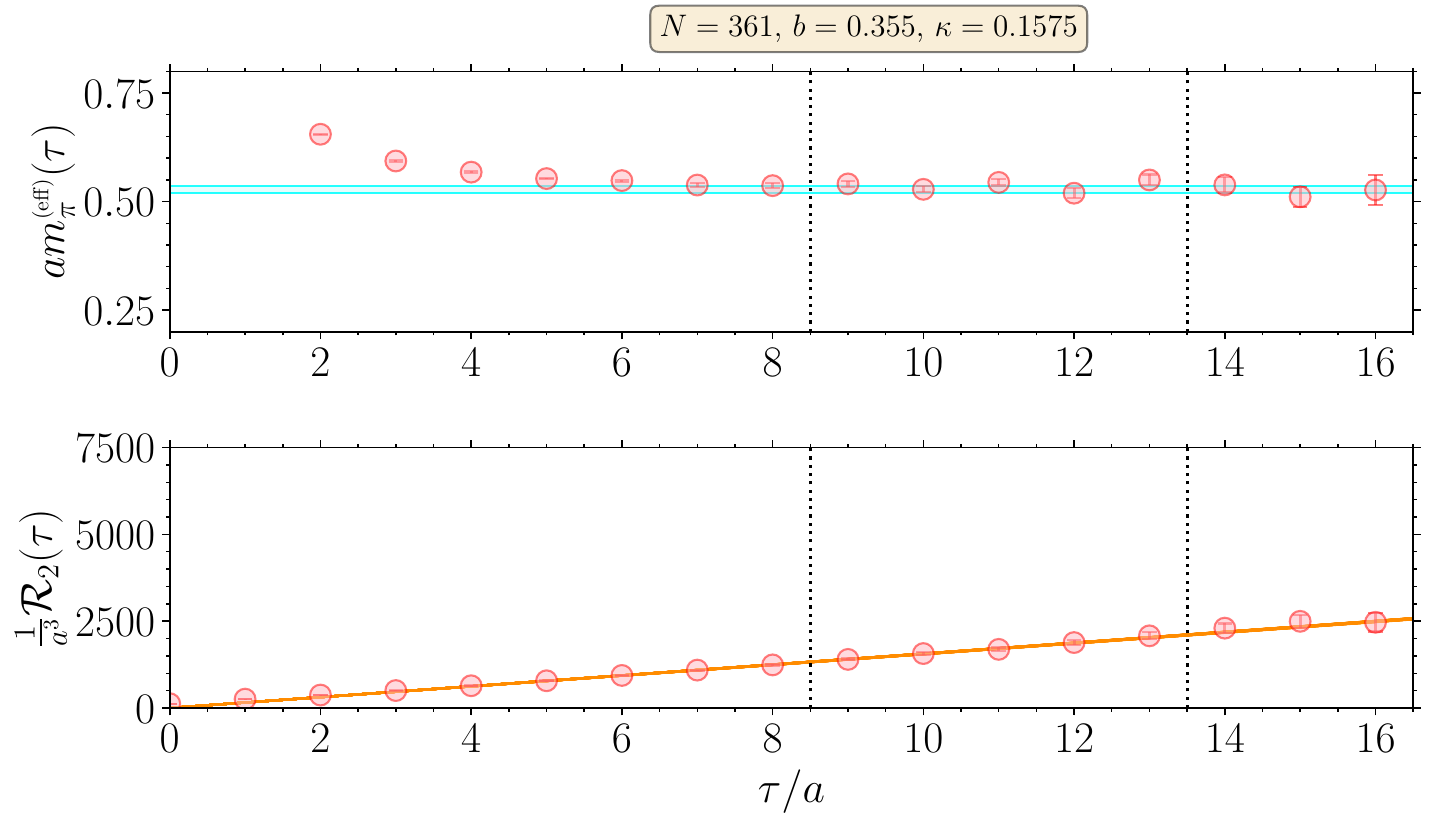}
\caption{Further examples of energy shift calculations.}
\label{fig:energy_shift_ex_additional}
\end{figure}

Figure~\ref{fig:energy_shift_ex_additional} shows additional examples of energy shift calculations. The top panel of Fig.~\ref{fig:energy_shift_ex_additional} has roughly the same lattice spacing and pion mass values as in Fig.~\ref{fig:energy_shift_ex}, and differs in the effective volume. The central panel of Fig.~\ref{fig:energy_shift_ex_additional} has the same lattice spacing and roughly the same effective volume as in Fig.~\ref{fig:energy_shift_ex}, and differs in the pion mass. Finally, the top and bottom panels of Figure~\ref{fig:energy_shift_ex_additional} have roughly the same effective volume and pion mass values, and differ in the lattice spacing.

\section{\texorpdfstring{Large-$N$}{Large-N} estimation of \texorpdfstring{$\bar{\ell}_3/N$}{l3/N}}\label{app:l3}

Using the data reported here and in Ref.~\cite{Bonanno:2025hzr} for the pion mass as a function of $b$ and $\kappa$, the results for the critical hopping $\kappa_c(b)$ and for the scalar renormalization constant $Z_{\sstrm{S}}(b)$, one is able to determine $m_\pi^2$ as a function of the renormalized quark mass $am_{\sstrm{R}} =Z^{-1}_{\sstrm{S}}\left(\frac{1}{2\kappa}-\frac{1}{2\kappa_c}\right)$, with $\kappa_c$ the critical Wilson hopping parameter. Since we have updated some values of $m_\pi$, we have also repeated chiral fits of $[am_\pi(b,\kappa)]^2$ versus $1/(2\kappa)$ at fixed $b$ to update the values of $\kappa_c(b)$ too. For completeness, we reported in Tab.~\ref{tab:mpi_vs_mquark} all the quantities entering the best fit discussed in this section.

\begin{figure}[!t]
\centering
\includegraphics[scale=0.62]{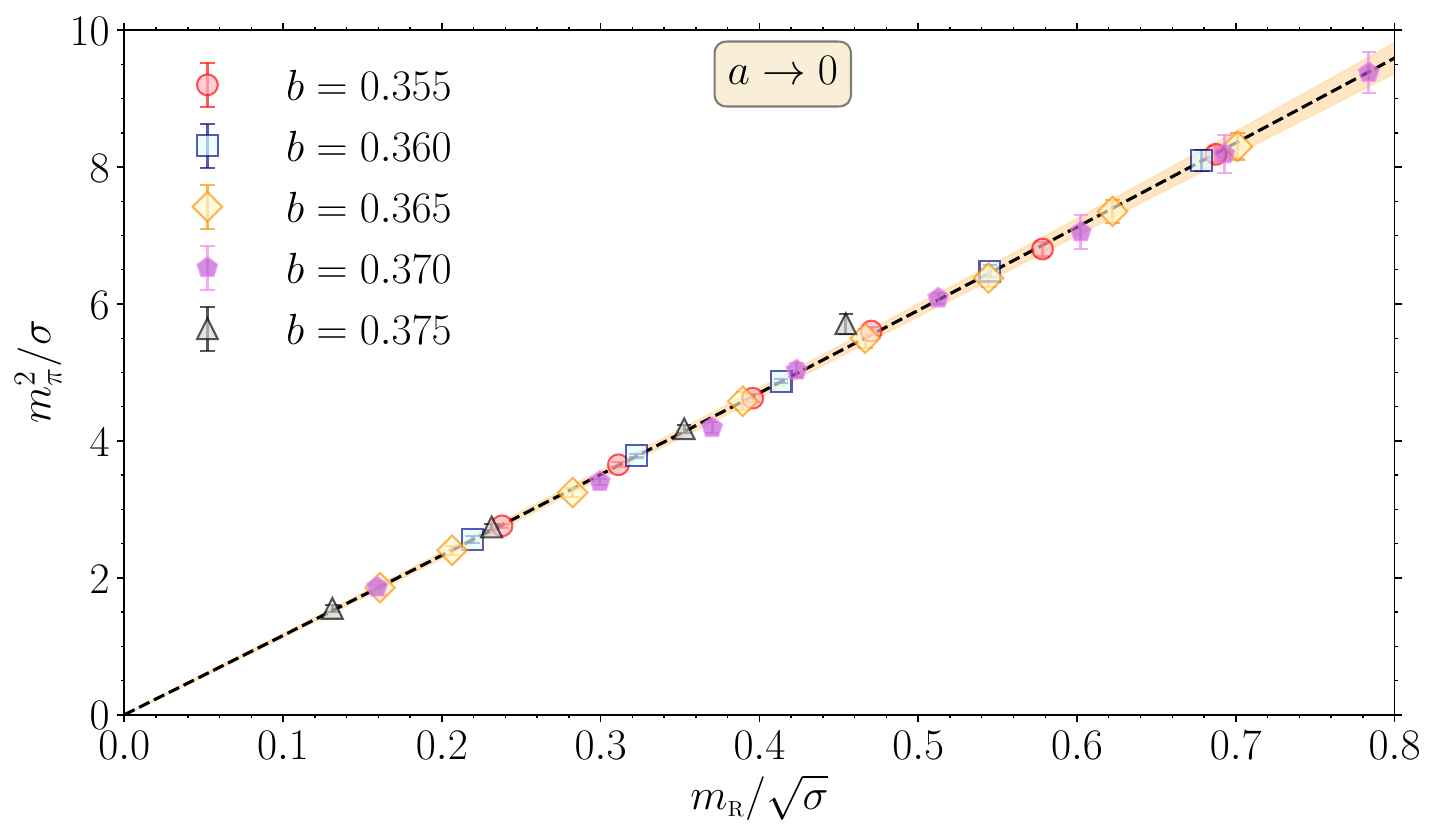}
\caption{Chiral behavior of the squared pion mass as a function of the renormalized quark mass in the continuum limit. The shaded line represents the best fit to Eq.~\eqref{eq:global_fit_func_mpi_vs_mR}, minus lattice artifact terms, which have also been subtracted from the data.}
\label{fig:mpi_vs_mR}
\end{figure}

Performing a four-parameter global chiral-continuum fit of the type:
\beq\label{eq:global_fit_func_mpi_vs_mR}
\frac{m_\pi^2}{\sigma} = K_0 \frac{m_{\sstrm{R}}}{\sqrt{\sigma}}\left[1 - \frac{K_1}{32\pi^2} \frac{K_0}{F^2/(N\sigma)}  \frac{m_{\sstrm{R}}}{\sqrt{\sigma}} + K_2 \, a \sqrt{\sigma}\right] + K_3 \, a \sqrt{\sigma},
\eeq
with $\frac{1}{\sqrt{N\sigma}}F=0.1262(34)$, we are able to provide an estimate of the large-$N$ value of $\bar{\ell}_3/N$ from the fit parameter $K_1$. The best fit procedure is illustrated in Fig.~\ref{fig:mpi_vs_mR}, and yields:
\beq
K_1 = \frac{\bar{\ell}_3}{N} = -0.022(22).
\eeq
The best fit parameter $K_0/2$ can be identified with $B_{\sstrm{R}}/\sqrt{\sigma} = \Sigma_{\sstrm{R}}/(F^2 \sqrt{\sigma})$ and turns out to be $K_0/2 = 5.76(15)$, which perfectly agrees with the determination $B_{\sstrm{R}}/\sqrt{\sigma}=5.58(26)$ reported in~\cite{Bonanno:2025hzr} (obtained performing individual chiral limits at fixed $b$ of the quark-mass dependence of the pion mass, and then extrapolating $B_{\sstrm{R}}$ towards the continuum limit).

\begin{figure}[!t]
\centering
\includegraphics[scale=0.62]{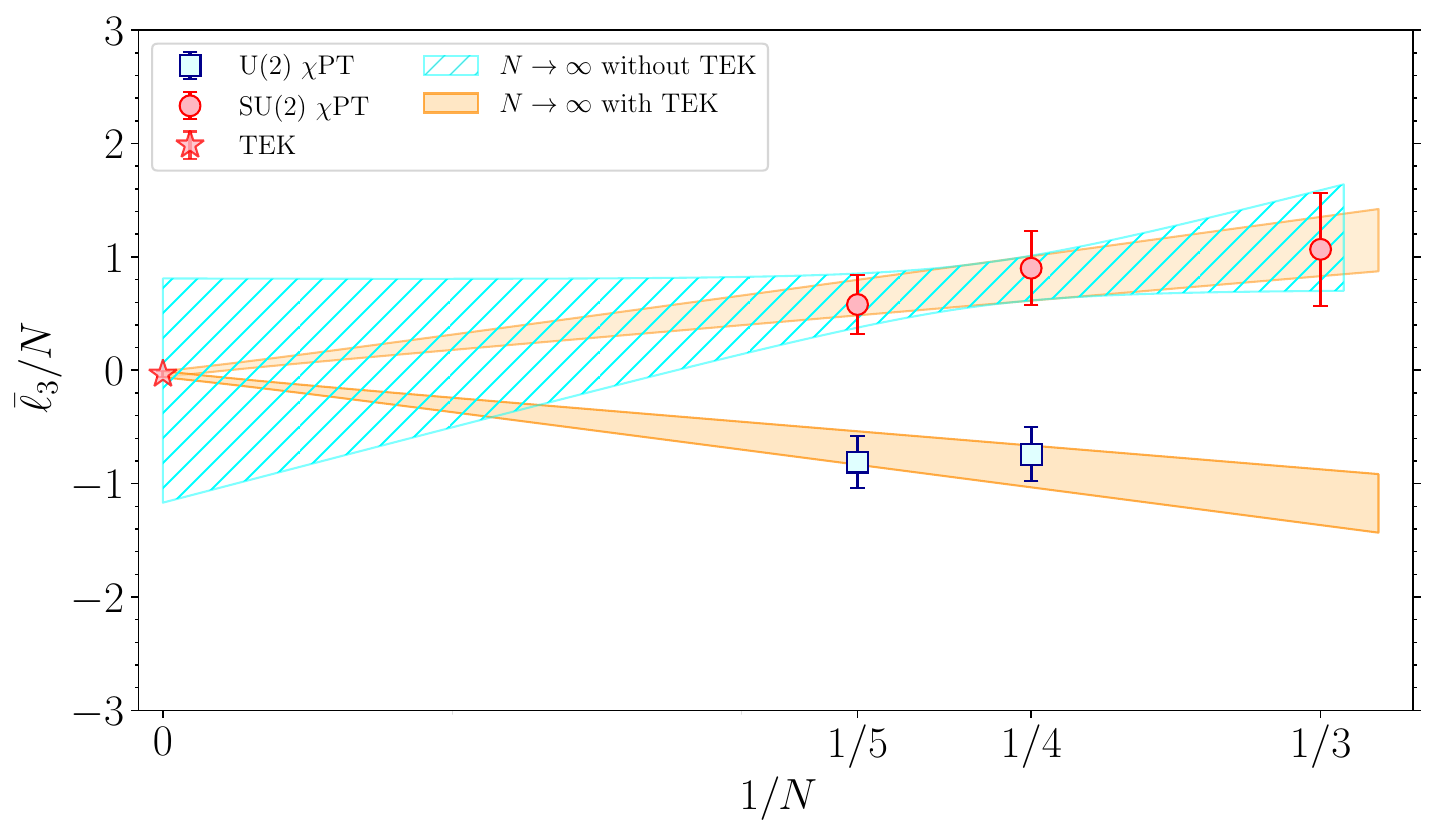}
\caption{Estimation of the $1/N$ expansion of the NLO LEC $\bar{\ell}_3/N$, obtained combining the large-$N$ TEK determination of this quantity with the $\Nf=2$ finite-$N$ results of~\cite{DeGrand:2023hzz}.}
\label{fig:l3_Ndep}
\end{figure}

Combining this result with the $\Nf=2$ finite-$N$ determinations of $\bar{\ell}_3$ of Ref.~\cite{DeGrand:2023hzz} we are also able to estimate the sub-leading $\mathcal{O}(1/N)$ term in the $1/N$ expansion of $\bar{\ell}_3/N$. This fit procedure, using data reported in Tab.~\ref{tab:l3_vs_N}, is illustrated in Fig.~\ref{fig:l3_Ndep}, and yields:
\beq
\frac{\bar{\ell}_3}{N} &=& -0.022(22) + 1.66(39)\frac{\Nf}{N} + \mathcal{O}\left(\frac{1}{N^2}\right), \qquad \qquad \SU(\Nf) \text{ $\chiPT$},\\
\label{eq:l3_U_Nf}
\frac{\bar{\ell}_3}{N} &=& -0.022(22) -1.66(37)\frac{\Nf}{N} + \mathcal{O}\left(\frac{1}{N^2}\right), \qquad \qquad \U(\Nf) \text{ $\chiPT$}.
\eeq
As it can be seen from Fig.~\ref{fig:l3_Ndep}, the TEK bound on the large-$N$ limit of $\bar{\ell}_3/N$ agrees with the one that would be obtained from $N<6$ data, but it is much more stringent.

\begin{table}[!t]
\begin{center}
\begin{tabular}{|c|c|c|}
\hline
&&\\[-1em]
$N$ & \makecell{$\bar{\ell}_3/N$\\SU(2) $\chiPT$} & \makecell{$\bar{\ell}_3/N$\\U(2) $\chiPT$}\\
\hline
3 & 1.07(50) &          \\
4 & 0.90(33) & -0.74(24)\\
5 & 0.58(26) & -0.81(23)\\
\hline
$\infty$ & \multicolumn{2}{c|}{-0.022(22)} \\
\hline
\end{tabular}
\end{center}
\caption{Continuum finite-$N$ lattice determinations of $\bar{\ell}_3/N$ for $\Nf=2$ according to $\SU(2)$ and $\mathrm{U}(2)$ $\chiPT$ fits obtained in Ref.~\cite{DeGrand:2023hzz}, along with the continuum large-$N$ TEK lattice determination presented in this article, obtained using data originally presented in Ref.~\cite{Bonanno:2025hzr}.}
\label{tab:l3_vs_N}
\end{table}

\begin{table}[!t]
\begin{center}
\begin{tabular}{|c|c|c|c|c|}
\hline
$b$ & $\kappa$ & $am_\pi$ & $\kappa_c(b)$ & $Z_{\sstrm{S}}(b)$ \\
\hline
0.355  &  0.1565  &  0.5798(64)  &  \multirow{6}{*}{0.163015(78)*}  &  \multirow{6}{*}{0.7287(23)} \\
0.355  &  0.1575  &  0.5277(67)* & & \\
0.355  &  0.1585  &  0.4821(36)* & & \\
0.355  &  0.1592  &  0.4367(43)* & & \\
0.355  &  0.1600  &  0.3904(28)* & & \\
0.355  &  0.1607  &  0.3407(28)* & & \\
\hline
0.360  &  0.1550  &  0.5081(68)  &  \multirow{5}{*}{0.160230(74)*} &  \multirow{5}{*}{0.7226(67)} \\
0.360  &  0.1560  &  0.4550(46)* &  & \\
0.360  &  0.1570  &  0.3947(16)* &  & \\
0.360  &  0.1577  &  0.3490(16)* &  & \\
0.360  &  0.1585  &  0.2887(40)* &  & \\
\hline
0.365  &  0.1535  &  0.4532(70)  &  \multirow{8}{*}{0.158075(63)*}  &  \multirow{8}{*}{0.732(11)} \\
0.365  &  0.1540  &  0.4267(65)  &  & \\
0.365  &  0.1545  &  0.3973(54)  &  & \\
0.365  &  0.1550  &  0.3698(63)  &  & \\
0.365  &  0.1555  &  0.3379(65)  &  & \\
0.365  &  0.1562  &  0.2845(37)  &  & \\
0.365  &  0.1567  &  0.2468(43)* &  & \\
0.365  &  0.1570  &  0.2179(78)  &  & \\
\hline
0.370  &  0.1520  &  0.4287(88)  &  \multirow{8}{*}{0.156418(38)*}  &  \multirow{8}{*}{0.743(14)} \\
0.370  &  0.1525  &  0.4002(86)  &  & \\
0.370  &  0.1530  &  0.3710(86)  &  & \\
0.370  &  0.1535  &  0.3458(29)  &  & \\
0.370  &  0.1540  &  0.3150(41)  &  & \\
0.370  &  0.1543  &  0.2859(33)* &  & \\
0.370  &  0.1547  &  0.2583(21)  &  & \\
0.370  &  0.1555  &  0.1939(26)  &  & \\
\hline
0.375  &  0.1527  &  0.2997(47)* &  \multirow{4}{*}{0.154958(40)*}  &  \multirow{4}{*}{0.755(18)} \\
0.375  &  0.1532  &  0.2551(21)  &  & \\
0.375  &  0.1538  &  0.2076(19)  &  & \\
0.375  &  0.1543  &  0.1579(31)  &  & \\
\hline
\end{tabular}
\end{center}
\caption{Results marked with * have been updated in this study, while the others come from Ref.~\cite{Bonanno:2025hzr}.}
\label{tab:mpi_vs_mquark}
\end{table}

\providecommand{\href}[2]{#2}\begingroup\raggedright\endgroup

\end{document}